\documentclass[11pt,preprintnumbers]{article}
\usepackage{jheppub}
\usepackage{bbm}
\usepackage{epsfig,bm}
\usepackage{graphics,graphicx,comment,subfigure,caption}
\usepackage{amsmath}
\usepackage{mdframed}
\usepackage{epigraph}
\usepackage{empheq}
\usepackage{framed}
\usepackage{slashed}

\newcommand{\be}{\begin{equation}}
\newcommand{\ee}{\end{equation}}

\def\bea{\begin{align}}
\def\ena{\end{align}}

\def\Tr{\mbox{Tr}}
\def\beqa{\begin{eqnarray}}
\def\enqa{\end{eqnarray}}

\def\a{\alpha}
\def\b{\beta}
\def\da{\dot{\alpha}}
\def\db{\dot{\beta}}

\newcommand{\bpsi}{\overline{\psi}\,}
\newcommand{\bsigma}{\overline{\sigma}\,}
\newcommand{\bepsilon}{\overline{\epsilon}\,}

\hypersetup{
    colorlinks=true,       
    linkcolor=blue,          
    citecolor=red,        
    filecolor=green,      
    urlcolor=magenta           
}

\begin{document}

\title{Bose-Fermi Degeneracies in Large $N$ Adjoint QCD}

\author[1,2]{G\"{o}k\c{c}e Ba\c{s}ar,}
\emailAdd{basar@tonic.physics.sunysb.edu} 
\author[3]{Aleksey Cherman,}
\emailAdd{cherman@physics.umn.edu} 
\author[4]{and David McGady}
\emailAdd{dmcgady@princeton.edu} 
\affiliation[1]{
Maryland Center for Fundamental Physics, Department of Physics,
University of Maryland, College Park, MD USA}
\affiliation[2]{
Department of Physics and Astronomy, Stony Brook University, Stony Brook, NY 11794, USA}
\affiliation[3]{Fine Theoretical Physics Institute, School of Physics and Astronomy, University of Minnesota, USA}
\affiliation[4]{Department of Physics, Princeton University, Princeton NJ 08544, USA}

\preprint{{\flushright PUPT-2466\\UMN-TH-3352/14\\FTPI-MINN-14/26\\}}

\abstract{
We analyze the large $N$ limit of adjoint QCD, an $SU(N)$ gauge theory with $N_f$ flavors of massless adjoint Majorana fermions,   compactified on $S^3 \times S^1$.  We focus on the weakly-coupled confining small-$S^3$ regime.   If the fermions are given periodic boundary conditions on $S^1$, we show that there are large cancellations between bosonic and fermionic contributions to the twisted partition function.  These cancellations follow a pattern previously seen in the context of misaligned supersymmetry, and lead to the absence of Hagedorn instabilities for any $S^1$ size $L$, even though the bosonic and fermionic densities of states both have Hagedorn growth. Adjoint QCD stays in the confining phase for any $L \sim N^0$, explaining how it is able to enjoy large $N$ volume independence for any $L$.  
The large $N$ boson-fermion cancellations take place in a setting where adjoint QCD is manifestly non-supersymmetric at any finite $N$, and are consistent with the recent conjecture that adjoint QCD has emergent fermionic symmetries in the large $N$ limit.   
}
\maketitle



\section{Introduction}
In this paper we explore adjoint QCD, an $SU(N)$ gauge theory with $N_f$ flavors of massless Majorana quarks in the adjoint representation of $SU(N)$.  Working in a weakly coupled and analytically tractable regime, we show that for any $N_f \ge 1$ there are large cancellations between bosonic and fermionic contributions to the $(-1)^F$-twisted partition function at large $N$.  The cancellations are so strong that when large $N$ adjoint QCD is compactified on a spatial circle of size $L$, with periodic boundary conditions for the fermions, it has no Hagedorn instabilities and stays in a confined phase for \emph{any} $L \sim N^0$, and enjoys large $N$ volume independence for any $L \sim N^0$.  

The weakly coupled regime used in our calculations opens up when the theory is compactified on $S^3 \times S^1$ and the $S^3$ radius is made small\cite{Sundborg:1999ue,Polyakov:2001af,Aharony:2003sx}. When the $S^1$ is large, the large $N$ theory can be shown to be in a confined phase, with the physical spectrum consisting of weakly coupled `hadron' states created by single-trace operators and an order $N^0$ free energy.   If the $S^1$ circle is spatial, with periodic boundary conditions for the fermions,  the Euclidean path integral computes the twisted partition function\cite{Unsal:2007fb}
\begin{align}
\tilde{Z}(L) = \Tr (-1)^F e^{-L H} =  \int d E\, \left[\rho_B(E) - \rho_F(E)\right]e^{-LE}
\label{eq:twistedZ}
\end{align}
where $\rho_{B,F}$ are the bosonic and fermionic densities of states and $L$ is the circumference of the $S^1$.  We verify that as a consequence of the Hagedorn phenomenon, both $\rho_B$ and $\rho_F$  grow exponentially in $E$.   In principle $\rho_B$ and $\rho_F$ might be expected to be quite different from each other.   Remarkably, we find that $\rho_B$ and $\rho_F$ have the same asymptotic behavior, with all exponentially-growing parts coinciding exactly for any $N_f\geq1$. Such a relation between the bosonic and fermionic densities of states leads to the dramatic consequence that adjoint QCD on $S^3 \times S^1$ \emph{does not} have a Hagedorn instability, and the theory stays in the confined phase for \emph{any} spatial circle size $L \sim N^0$ for any $N_f \ge 1$. This is due to the fact that \eqref{eq:twistedZ} involves $\rho_B-\rho_F$, in contrast to the thermal partition function, which involves $\rho_B +\rho_F$.  The boson-fermion degeneracies lead to strong cancellations in \eqref{eq:twistedZ}, and keep $\tilde Z(L)$ a smooth function of $L$ for any $L \sim N^0$.   Our results provide physical insight into the result of \cite{Unsal:2007fb}, which found that adjoint QCD on $S^3 \times S^1$ enjoys large $N$ volume independence for any $L$.
 
The observation of degeneracies between bosonic and fermionic spectra normally suggests that the theory has a fermionic symmetry.  But at any finite $N$, adjoint QCD on $S^3 \times S^1$ is not supersymmetric.    The $S^3$ curvature breaks the flat-space $\mathcal{N}=1$ supersymmetry of the $N_f=1$  theory, while if $N_f>1$ the theory has $2(N^2-1)$ bosonic and $2N_f (N^2 -1)$ fermionic degrees of freedom at the microscopic level, and hence cannot be supersymmetric in any conventional sense even in flat space.  Since the degeneracies we observe appear in the large $N$ limit, our results are consistent with the conjecture posed in  \cite{Basar:2013sza} that adjoint QCD should have an \emph{emergent} fermionic symmetry in the large $N$ limit even away from $N_f=1$ if the theory enjoys volume independence.  Emergent fermionic symmetries in the large $N$ limit of otherwise non-supersymmetric theories do not contradict the Coleman-Mandula and Haag-Lopuszanski-Sohnius theorems, since the $S$-matrix elements of physical states vanish in the large $N$ limit.

The paper is organized as follows.  In Section \ref{sec:Tension} we review some relevant properties of adjoint QCD, and summarize the arguments of \cite{Basar:2013sza} concerning Hagedorn instabilities and large $N$ volume independence  which motivated our search for spectral degeneracies in adjoint QCD.   In Section \ref{sec:PartitionFunction} we describe the calculation of the twisted and thermal partition functions for adjoint QCD in the large $N$ limit on $S^3 \times S^1$, using the technology of  \cite{Sundborg:1999ue,Polyakov:2001af,Aharony:2003sx}.  Section \ref{sec:InstabilitiesAndDisappearance} is the key part of the paper, and describes the behavior of the twisted and thermal densities of states which are relevant for spatial and thermal compactifications respectively. Figure \ref{fig:singularityPlot} gives a visual summary of our story.  Thermally-compactified adjoint QCD has Hagedorn instabilities, as shown in Section \ref{sec:ThermalInstability}, but there are no Hagedorn instabilities for spatial compactification as shown in Section \ref{sec:SpatialStability}.  We compute the twisted Casimir energy in adjoint QCD at large $N$ and show that it vanishes in Section \ref{sec:Casimir}, while  Section \ref{sec:Misaligned} comments on the connections between our results and misaligned supersymmetry.  Finally, in Section \ref{sec:Symmetries}, we make some remarks on the relation of our findings to the underlying symmetries of adjoint QCD, and conclude in Section \ref{sec:Conclusions}.
\section{Properties of large $N$ adjoint QCD}
\label{sec:Tension}
In this section we briefly review two properties of large $N$ gauge theories --- and in particular of adjoint QCD --- which play a key role in the rest of our analysis.  These properties are the presence of Hagedorn instabilities in generic confining large $N$ gauge theories, and the phenomenon of large $N$ volume independence, which is special to adjoint QCD.   The tension between Hagedorn instabilities and volume independence motivate our study of adjoint QCD on $S^3 \times S^1$.

\subsection{Hagedorn instability}
Large $N$ gauge theories with a confinement scale $\Lambda_c$ are believed to have a density of states $\rho(E)$ with a Hagedorn scaling \cite{Hagedorn:1965st}
\begin{align}
\rho(E \gg \Lambda_c) \to e^{\beta_H E}, \qquad \beta_H \sim \Lambda_c^{-1}
\label{eq:HagedornDefinition}
\end{align}
A heuristic argument for this relation is that large $N$ theories have an infinite number of stable hadronic states, and highly-excited states can be thought of as excitations of confining strings, see e.g. \cite{Polchinski:1998rq}.  Relativistic string theories famously have Hagedorn densities of states, motivating \eqref{eq:HagedornDefinition}. A more rigorous argument in favor of \eqref{eq:HagedornDefinition} based directly on the known properties of large $N$ gauge theories was recently given in \cite{Cohen:2009wq,Cohen:2011yx}. 

If such a theory is compactified on $M \times S^1_{\beta}$, where $S^1_{\beta}$ is a thermal circle, then the associated partition function can be written as
\begin{align}
Z(\beta) = \Tr\, e^{-\beta H} =  \int d E \left[\rho_B(E) + \rho_F(E)\right]e^{-\beta E}
\end{align}
with $\rho_{B, F}$ being the bosonic and fermionic densities of states respectively.  If $\rho_B +\rho_F = \rho$ satisfies \eqref{eq:HagedornDefinition}, then the sum over states in $Z(\beta)$ will diverge for $\beta \le \beta_H$.  This is known as a Hagedorn instability.  Consequently, it is believed that all confining large $N$ theories undergo a deconfinement phase transition at some inverse temperature $\beta_d \ge \beta_H$.

\subsection{Large $N$ volume independence}
Consider a confining gauge theory with one or more directions compactified on a spatial torus $T$ with periodic boundary conditions for fermions, and suppose the theory is in the confining phase.  In general, connected correlation functions of single-trace color singlet operators will depend on the volume of $T$, with the dependence taking the form $e^{-L \Lambda}$ where $\Lambda$ is the mass gap and $L \sim N^0$ is the scale of the volume\footnote{The restriction to $L\sim N^0$ is important, since in general volume \emph{dependence} is expected to set in once $L \sim N^{-1}$, with e.g. possible chiral phase transitions at $L \sim 1/(N\Lambda)$ where $\Lambda$ is the strong scale.  The restriction to toroidal compactifications is also important, since on e.g. $S^3_R \times S^1_L$ the physics depends on $R$ even at large $N$, in contrast to what sometimes happens to the dependence on $L$.  }. Large $N$ volume independence is the statement that in the `t Hooft large $N$ limit,  the connected correlation functions of topologically trivial single-trace operators do not depend on $L$, provided  center symmetry and translation invariance are not broken \cite{Eguchi:1982nm,Bhanot:1982sh,Narayanan:2003fc,Cohen:2004cd,Kovtun:2007py,Unsal:2010qh}\footnote{There is a simple heuristic picture behind the phenomenon of large $N$ volume independence.  The way a given hadron knows that it is a periodic box is to interact with the  `image' hadrons introduced by the boundary conditions on the walls.  If we take an `t Hooft large $N$ limit, with $N\to \infty$ with all physical scales fixed, then the interactions between hadrons become $1/N$ suppressed, and the finite volume effects must disappear at leading order in the $1/N$ expansion.  So as long as a large $N$ theory is in its confining phase, it will enjoy volume independence for toroidal compactifications.  }.  
 Volume independence implies that the connected parts of $n\ge1$-point correlation functions of single-trace topologically-trivial operators are $L$-independent up to $1/N$ corrections.   For zero point-functions such as $\log Z$ (the free energy), volume independence forces their $\mathcal{O}(N^2)$ parts to be volume independent.  Of course, in the confining phase, where center symmetry is unbroken and volume independence is valid, $\log Z$ is $\mathcal{O}(N^0)$.   Hence the validity of volume independence for $L \in [L_{\rm min}, \infty)$ implies that a theory must not have any Hagedorn instabilities for $L \in [L_{\rm min}, \infty)$, since these would drive the appearance of an $\mathcal{O}(N^2)$ volume-dependent part in $\log Z$.

Recently, convincing numerical and analytic evidence\cite{Cossu:2009sq,Bedaque:2009md,Bringoltz:2009mi,Bringoltz:2009kb,Hietanen:2009ex,Poppitz:2009fm,Azeyanagi:2010ne,Poppitz:2010bt,Hietanen:2010fx,Dorigoni:2010jv,Catterall:2010gx,Bringoltz:2011by,Armoni:2011dw,GonzalezArroyo:2012st,Gonzalez-Arroyo:2013bta,Gonzalez-Arroyo:2013gpa} has appeared that adjoint QCD with massless quarks is special in the sense that, when compactified on $M\times S^1_L$, it enjoys large $N$ volume independence  for \emph{any} circle size $L \sim N^0$\cite{Kovtun:2007py}
 so long as the circle is a \emph{spatial} one, with periodic boundary conditions for fermions.  That is, in adjoint QCD, large $N$ volume independence is believed to hold for $L \in (0, \infty)$ for any $N_f \in [1, 5.5)$.~\footnote{When $N_f < 5.5$, adjoint QCD  is asymptotically-free and has a strong scale $\Lambda$ as determined from the IR Landau pole in the one-loop beta function.  For $N_f < 4$ adjoint QCD on $\mathbb{R}^4$ is believed to develop a mass gap of order $\Lambda$.  If $5.5>N_f \gtrsim 4$, it is believed that adjoint QCD on $\mathbb{R}^4$ flows to a conformal fixed point in IR, and for $N_f=5$ this fixed point can be seen in the two-loop beta function, and occurs at weak coupling.}

\subsection{The tension}
Volume independence for any $L $ implies the absence of phase transitions as a function of $L$.   As a result, one might worry that large $N$ volume independence for any $L$ is not consistent with the well-established existence of Hagedorn instabilities at $L_H \sim \Lambda_c^{-1}$ in confining theories.  Indeed, in many theories there truly is a clash between volume independence and the Hagedorn instability, which is resolved by the failure of volume independence at $\beta = \beta_d$\cite{Cohen:2004cd,ShifmanPrivate:2012}.  From a modern perspective, this gives a simple heuristic explanation for the failure of the original large $N$ volume independence proposal of Eguchi and Kawai in the context of pure Yang-Mills theory\cite{Eguchi:1982nm,Bhanot:1982sh}.   However, adjoint QCD does not necessarily suffer from this issue\cite{Basar:2013sza}.  To see this, recall that the modern formulation of large $N$ volume independence is a statement about the sensitivity of observables to the size of spatial circles\cite{Kovtun:2007py}. The Euclidean path integral for a theory compactified on a spatial circle computes the twisted partition function, $\tilde Z(L)$, defined in \eqref{eq:twistedZ}; it does not compute 
 the thermal partition function $Z(\beta)$.   The twisted and thermal partition functions are sharply different in theories with bosonic and fermionic states of similar energies.  This is the case in $SU(N)$ adjoint QCD with massless fermions.  In contrast, in QCD with $N_f$ fundamental fermions, with even $N$ there  are no fermionic states at all, while for odd $N$ the only fermionic states are baryons, which become parametrically heavy in the large $N$ limit.
The general statement is that the twisted and thermal partition functions are qualitatively similar for $\beta \sim L \sim N^0$ for large $N$ gauge theories with complex-representation fermions, but they are very different in theories with light adjoint fermions. 

The relevance of $\tilde{Z}(L)$ rather than $Z(\beta)$ means that the tension between volume independence and Hagedorn instabilities would be relieved if the exponentially-growing parts of $\rho_B$ and $\rho_F$ were the same, leading to sufficient cancellations in \eqref{eq:twistedZ} to avoid Hagedorn instabilities.   Supersymmetry would of course be sufficient to drive such cancellations, since in flat space the twisted partition function of a supersymmetric QFT is the Witten index, which is trivially volume-independent.  

However, adjoint QCD is not supersymmetric for generic $N_f$, so it is not a priori obvious why one should expect sufficient cancellations in the twisted partition function to avoid Hagedorn instabilities.   In this paper we show that the necessary cancellations do indeed happen in adjoint QCD on $S^3\times S^1$ for any $N_f \ge 1$.  Since our results involve degeneracies between the energies of an infinite number of bosonic and fermionic states, it appears to call for the presence of emergent fermionic symmetries in large $N$ adjoint QCD.  

\subsection{Utility of $S^3 \times S^1$ compactifications}
Both volume independence and Hagedorn instabilities are usually strong coupling phenomena, which makes their  interplay difficult to explore analytically.  In this paper we discuss volume independence and Hagedorn instabilities in adjoint QCD on $S^3_R \times S^1_{\beta}$ and $S^3_R \times S^1_{L}$, using methods developed in \cite{Aharony:2003sx,Sundborg:1999ue,Unsal:2007fb}.  
The reason this setting is interesting is that if $N_f < 5.5$, then the 't Hooft coupling $\lambda(R) \to 0$ as $\Lambda R \to 0$, where $\Lambda$ is the strong scale.  Hence the theory becomes weakly coupled and analytically calculable for \emph{any} $L$ or $\beta$.\footnote{Our results also apply if $N_f > 5.5$, when the theory becomes IR-free, with a Landau pole $\Lambda$ for the coupling in the UV.  In this regime we can maintain weak coupling by setting  $R\Lambda \gg 1$. }  At the same time, the $\Lambda R \ll 1$ theory is confining with a mass gap of order $1/R$, with the realization of center symmetry serving as an order parameter for confinement.  As we will verify using the techniques of \cite{Aharony:2003sx,Sundborg:1999ue}, the presence of a Hagedorn density of states in adjoint QCD can be shown by direct calculation so long as $\Lambda R \ll 1$. Consequently, the $R \Lambda \ll 1$ limit gives us a regime where Hagedorn phenomena, center symmetry realizations and large $N$ volume independence can all be explored simultaneously at weak coupling. 

The presence of $S^3$ curvature couplings explicitly breaks the flat-space supersymmetry of the $N_f=1$ $SU(N)$ theory, while $N_f>1$ adjoint QCD is not supersymmetric even in flat space.  So one might worry that on $S_R^3 \times S_L^1$, volume independence would be doomed both with $N_f=1$ and $N_f >1$.   However, some time ago, it was shown by \"Unsal\cite{Unsal:2007fb} that in adjoint QCD center symmetry is always unbroken on $S^3_R \times S^1_L$ for any $N_f \ge 1$, and hence large $N$ volume independence must hold for any $N_f \ge 1$.\footnote{See also \cite{Hollowood:2009sy} for a discussion of the fate of volume independence in this setting when a quark mass is turned on.}  We illuminate the physics of this result by explicitly showing that there are no Hagedorn instabilities  any $N_f \ge 1$ for any $L \sim N^0$ in the spatially-compactified theory.  On the other hand, we show that there \emph{are} Hagedorn instabilities for thermal compactification with $\beta \sim 1/R$.  The spatially-compactified theory with $N_f \ge 1$ avoids Hagedorn instabilities due to large cancellations between bosonic and fermionic densities of states, as was advocated on general grounds in \cite{Basar:2013sza}.

Before diving into the analysis, we make a remark on the global symmetries of adjoint QCD.  Since the $N_f$ Majorana fermions are in a real representation of the gauge group, the theory has a classical $U(N_f)$ flavor symmetry.  The overall $U(1) \subset U(N_f)$ is anomalous, and on $\mathbb{R}^3 \times S^1$ it is believed that $SU(N_f)$ is spontaneously broken to $SO(N_f)$ by a chiral condensate when the $S^1$ is large.\footnote{See e.g.~\cite{Unsal:2007vu,Unsal:2007jx,Unsal:2008eg,Nishimura:2009me,Anber:2011gn,Misumi:2014raa,Misumi:2014jua} for studies of confinement and chiral symmetry breaking in adjoint QCD in the volume-\emph{dependent} weakly coupled regime which opens up for spatial circle compactification if $N L\Lambda \ll 1$. See also \cite{Shifman:2013yca} for a recent overview of some properties of adjoint QCD. }  The situation is quite different on $S^3_R \times S^1_L$, since the chiral symmetry realization depends on $R \Lambda$.   For small $R \Lambda$, where the theory is weakly coupled for any $L \sim N^0$, the $SU(N_f)$ chiral symmetry is not spontaneously broken, and the curvature couplings induce a chirally-symmetric mass gap for the fermions\cite{Unsal:2007fb}.  The small $R \Lambda$ regime is an example of a setting where confinement and chiral symmetry breaking are not entangled with each other.   These remarks will be important in Section~\ref{sec:Symmetries}.

\section{Large $N$ partition functions on $S^3 \times S^1$}
\label{sec:PartitionFunction}
When $R \Lambda \ll 1$, large $N$ adjoint QCD is a nearly free quantum theory with an infinite number of degrees of freedom.  Since all of the fields in the theory transform in the adjoint of the gauge group, in the $\lambda \to 0$ limit, each one of these degrees of freedom can be represented by $N \times N$ matrix harmonic oscillators, which transform as color-adjoints.  The frequency of each oscillator is of order $1/R$.   On a compact space, the Gauss law constraint, which applies no matter how small $R\Lambda$ becomes, implies that the only states which can contribute to a partition function must be color singlets.\footnote{The heuristic reason for this is that if one tries to put a source for color charge on a three-sphere there is no place for the color-flux lines to end.  In flat space, in contrast, the flux lines have the option of `ending' at the boundary at infinity.}    Hence all the matrix oscillators have to occur inside color traces, and a typical state looks something  like
\begin{align}
\Tr[B^{\dag}_{43} B^{\dag}_2 B_2^{\dag} B^{\dag}_{17} F^{\dag}_{9}] |0\rangle
\end{align}
where $B^{\dag}_i, F^{\dag}_i$ are bosonic and fermionic oscillator creation operators, respectively, with spin and flavor indices suppressed for simplicity.  

We will confine our attention to the behavior of adjoint QCD in the 't Hooft large $N$ limit.  This means sending $N$ to infinity while fixing (i) $N_f$, (ii) 't Hooft coupling $\lambda = g^2 N$, (iii) $S^3$ radius $R$, and (iv) the circle sizes $L$ or $\beta$.  Thanks to Boltzmann suppression factors, the last condition means that the only states that can contribute significantly to the partition function  have energies of order $N^0$.    When $R\Lambda \ll 1$, the energy of a state created by an a single-trace operator is directly proportional to the number of oscillators entering the trace.  Thus by working in the 't Hooft large $N$ limit defined by the conditions (i)-(iv) we are justified in only considering states created by $N^0$ oscillators.  This is a major simplification, because it means that the space of multi-trace states is the Fock space of single-trace states.\footnote{If the number of oscillators entering a single-trace operator scales with $N$ there are algebraic relations between the single-trace operator and linear combinations of multi-trace operators, making the state counting much more complicated.  These relations can be thought of as representing interactions between hadrons, which are $1/N$ suppressed for light states but may be unsuppressed for heavy states, as is well known from studies of large N baryons\cite{Witten:1979kh}.  These subtleties become important at finite $N$, and also become important if we consider non-'t Hooft large limits where we allow $L$ to scale as $1/N$.}

Combinatorially, the partition function of a system is a generating function which counts the number of states of each energy.  In the rest of this section, we review the technology\cite{Sundborg:1999ue,Polyakov:2001af,Aharony:2003sx} that lets one directly count the states in the large $N$ limit provided that $R \Lambda \ll 1$.  First, we recall how to count the independent $B_i$ and $F_i$ operators, taking into account gauge freedom and the equations of motion.  Then we count the single-trace and multi-trace color-singlet states.  All this is already known from \cite{Sundborg:1999ue,Polyakov:2001af,Aharony:2003sx}, but we repeat it here to keep the presentation self-contained.  At the end of the section we obtain exact expressions for the thermal and twisted partition function of adjoint QCD at large $N$ in the weakly coupled small $R$ limit.

\subsection{Single particle partition functions}
Adjoint QCD has a gauge field $A_{\mu}$ and fermion fields $\psi_a, a=1, \ldots, N_f$.   To build up a single-trace state, one can put together states composed of (a) various combinations of derivatives acting on $A_{\mu}$, as well as (b) various combinations of derivatives acting on $\psi$.  It is convenient to define generating functions $z_V$ and $z_F$ which count the number of independent color-adjoint states of type (a) and type (b) respectively. Following tradition we will call $z_V$ and $z_F$ ``single particle'' partition functions, though we emphasize that they are not the generating functions for the \emph{physical} single-particle states of a non-Abelian gauge theory. The state-operator correspondence maps the energies associated with these states, $E_{V,F}$,  to their classical scaling dimensions, $\Delta_{E,F}$,  as $E_{V,F} = \Delta_{F,V}/R$ on $S^3_{R} \times S^1_{L\, \mathrm{or}\, \beta}$ in the $R\Lambda \ll 1$ limit, and provides an easy way to calculate the single particle partition functions as
\begin{align}
z_F(q) &= \sum_{\Delta_F} d_{\Delta_F} q^{\Delta_F}\\
z_V (q) &= \sum_{\Delta_V} d_{\Delta_V} q^{\Delta_V}. 
\end{align}
Here $d_{\Delta_{F,V}}$ denotes the degeneracy of the operator with dimension $\Delta_{F,V}$ and $q =e^{-\beta/R}$ or $q=e^{-L/R}$ depending on whether we consider thermal or spatial compactification respectively. Explicitly counting the operators by taking into account the equations of motion and gauge constraints, one obtains \cite{Sundborg:1999ue,Aharony:2003sx,Polyakov:2001af}
\begin{align} 
z_F(q) &= {4 q^{3 \over2} \over (1-q)^3}  \label{letterZ} \\ \nonumber
z_V(q) &= \frac{6 q^2-2q^3}{(1-q)^3}\, . 
\end{align}
See Appendix~\ref{AppendixLetter} for a review of the derivations of these functions. Notably, these single particle partition functions have simple properties under the $T$-reflection symmetry $\beta \to -\beta$ introduced in \cite{Basar:2014mha}:
\begin{align}
z_F(1/q) &= - z_F(q) \label{Reflection0} \\\nonumber
1-z_V(1/q)& = - \big(1-z_V(q)\big)\, .  
\end{align}
These $T$-reflection properties are very useful for obtaining analytic expressions for the Hagedorn temperatures of the theory, as well as for being able to write the full partition functions in terms of elliptic functions.
\subsection{Twisted and thermal partition functions of adjoint QCD}
We now write down the twisted and thermal partition functions.  To get some intuition on the physics, note that at large $N$ we expect single-trace states to make the dominant contribution in the confined phase.  A rough estimate of the contribution to the partition function from e.g. the gauge fields is 
\begin{align}
Z_{\rm ST,\, naive} &= \sum_{k=1}^{\infty} \frac{1}{k} \left[z_V(q)\right]^k = - \log [1-z_V(q)]
\label{eq:ToyModelZ}
\end{align}
This naive estimate counts single-trace operators made with $k$ oscillators with a factor of $1/k$ to account for the cyclicity of the trace.    The counting entering this  estimate does not correctly deal with the combinatorics of repetitions of oscillators inside a single-trace, and multi-particle contributions are neglected. Both of these omissions lead to an undercounting of the states.  Nevertheless, the naive estimate above manages to capture the leading asymptotics of the state degeneracies, which control e.g. the Hagedorn temperature, so it is useful to keep it in mind in what follows.   

As shown in \cite{Sundborg:1999ue,Polyakov:2001af,Aharony:2003sx} the proper way to count the single-trace states with the correct weight for repetitions involves the use of Polya theory.  The result is
\begin{align}
Z_{\rm ST}[q] &= -\sum_{m=1}^{\infty} \frac{\varphi(m)}{m} \log \left[1-z_V(q^m) +(-1)^m N_f z_F(q^m)\right] \, , \\
\tilde{Z}_{\rm ST}[q] &= -\sum_{m=1}^{\infty} \frac{\varphi(m)}{m} \log \left[1-z_V(q^m) +N_f z_F(q^m)\right] \, .
\label{eq:ZST}
\end{align}
Here, $\varphi(m)$, the Euler totient function, is the number of positive integers less than or equal to, and relatively prime to $m$. In the 't Hooft large $N$ limit, the full confining-phase partition function can be obtained from the one above by including contributions from states involving an arbitrary number of particles.   The full large $N$ partition function can be written as\cite{Aharony:2003sx} \footnote{This construction, and its generalizations to finite $N$, is sometimes referred to as the `plethystic exponential', popularized in the physics literature in \cite{Benvenuti:2006qr,Feng:2007ur}.}
\begin{align}
\log Z[q] = \sum_{k=1}^{\infty} \frac{Z_{\rm ST}[q^k]}{k}.
\end{align}
Euler's formula, $\sum_{k|n} \varphi(k)=n$, then implies 
\begin{align}
\log Z[q] &= - \sum_{k=1}^{\infty} \log \left(1 - z_V(q^k)+(-1)^k N_f z_F(q^k)\right) 
\label{eq:FullPartitionThermal}\\
\log \tilde{Z}[q] &= - \sum_{k=1}^{\infty} \log \left(1 - z_V(q^k)+N_f z_F(q^k)\right)
\label{eq:FullPartitionSpatial}
\end{align}
Note that these expressions are \emph{only} correct at large $N$.  At finite $N$ (or in non-'t Hooft large $N$ limits) there are relations between e.g. single-traces with $\gtrsim N$ oscillators and multi-trace states, and such relations are ignored in the derivation leading to the above result.  

Before giving more explicit expressions for the partition functions, we make an important observation regarding the fermionic contributions to the single-trace and full partition functions. Due to the $q^{3/2}$ term in the fermionic single particle partition function, the fermions contribute to the expansions of the single-trace and full partition functions as half integer powers of $q$. Furthermore from Eqs. \eqref{letterZ}, \eqref{eq:FullPartitionThermal} and \eqref{eq:FullPartitionSpatial} we see that going from the thermal to the twisted compactification amounts to flipping the sign of the coefficients of the half integer powers of $q$, so that 
\begin{align}
Z &=\sum_{n=0}^\infty c_n q^n + \sum_{n=0}^\infty c_{n+1/2} \,q^{n+1/2}\\ 
\tilde Z &=\sum_{n=0}^\infty c_n q^n - \sum_{n=0}^\infty c_{n+1/2} \,q^{n+1/2}.
\end{align}
So as expected, the difference between the twisted and the thermal partition functions is that all the fermionic degeneracy factors (i.e. coefficients of the half integer powers of $q$) enter with a negative sign to the twisted partition function.  It is convenient to make the substitution $Q \equiv q^{1/2}$, so that the partition functions are power series expansion in $Q$ with the even and odd powers of corresponding to bosons and fermions, respectively.  

We now give give the expressions for the full partition functions in a more useful form. With the explicit single particle partition functions in Eq.~(\ref{letterZ}), the large $N$ pure YM partition function is
\begin{align}
Z_{YM}(q) = \tilde{Z}_{YM}(q) = \prod_{k=1}^{\infty} \frac{(1-q^k)^3}{(1+q^k)(c - q^k)(c^{-1}-q^k)} 
\label{eq:YMZ}
\end{align}
where $c = 2 +\sqrt{3}$~\footnote{The constant $c=2+\sqrt{3}$ appearing in the pure YM expression is a solution of \eqref{polynomial} for the variable $q=Q^2$ with $N_f=0$, along with $-1$ and $1/c$. }.  For pure YM, there is no difference between twisted and thermal partition functions by definition, since there are no fermionic states. Defining 
\begin{align}
1-z_V(Q^2)- N_F z_F(Q^2)={Q^6-3Q^4-4N_fQ^3-3Q^2+1\over(1-Q^2)^3} =: {P(Q)\over(1-Q^2)^3}
\end{align}
 with $N_f$ massless adjoint fermions, the thermal partition function is
\begin{align}
Z_{\rm QCD[Adj]} (Q) =\prod_{k=1}^{\infty}\frac{(1-Q^{2k})^3}{\prod_{i=1}^6(r_i +(-Q)^k)}, 
\end{align}
where $Q = q^{1/2}=e^{-\beta/2R}$ and $r_i$ with $i=1, 2, \ldots, 6$ are the six solutions of the equation 
\begin{align}
P(Q)=Q^6-3 Q^4-4 N_f Q^3-3 Q^2+1=0
\label{polynomial}
\end{align}
Note that, due to the $Q\rightarrow1/Q$ $T$-reflection symmetry of the equation \eqref{polynomial}, the roots of $P(Q)$ come in reciprocal pairs. Organizing the roots as $r_{4,5,6}\equiv1/r_{1,2,3}$, we obtain
  \begin{align}
Z_{\rm QCD[Adj]} (Q) =\prod_{k=1}^{\infty}\prod_{i=1}^3 \frac{(1-Q^{2k})}{\left(1+r_i(-Q)^k\right)\left(1+r_i^{-1}(-Q)^k\right)} 
\label{eq:ZNfThermal}
\end{align}
The exact expressions for the roots $r_i$ are given in Appendix \ref{Ap:Roots}. 

As discussed above, the twisted partition function can be obtained by taking $Q\rightarrow-Q$ in the thermal partition function, and it is given as
\begin{align}
\tilde Z_{\rm QCD[Adj]} (Q) =\prod_{k=1}^{\infty}\prod_{i=1}^3 \frac{(1-Q^{2k})}{ (1+ r_iQ^k)(1 +r_i^{-1} Q^k)} 
\label{eq:ZNfTwisted}
\end{align}

For completeness, note that the twisted partition function can also be written in terms of elliptic functions as 
\begin{align}
\tilde Z_{\rm QCD[Adj]}(L) &=\eta^3\left ( {i L \over 4 \pi R} \right)\eta^3\left ({i L \over 2 \pi R} \right)\prod_{i=1}^3 \left[ {r_i^{1/2}+r_i^{-1/2} \over \vartheta_2\left(\nu_i| e^{-{L\over 4 R}} \right)   } \right].
\label{eq:Zelliptic}
\end{align}
where $e^{2i\nu_i}\equiv r_i$, and the derivation is given in Appendix \ref{Ap:Elliptic}.
Here $\eta(\tau)=e^{i \pi \tau\over 12}\prod_{n=1}^\infty(1-e^{2i\pi \tau n})$ is the Dedekind eta function and
\begin{align}
\vartheta_2(u | e^{i\pi\tau})= \sum _{n=-\infty}^{\infty } e^{i (n+{1/2})^2\,\pi \tau}  e^{ (2 n+1) i u}\,,
\end{align} 
with $Q  = e^{-{L \over 2R}} =: e^{2 i \pi  \tau}$.

\section{Instabilities and their disappearance}
\label{sec:InstabilitiesAndDisappearance}

Equipped with the exact formulas for the partition functions, we now discuss Hagedorn instabilities.  In this section we show that the bosonic and fermionic states have identical asymptotics for $N_f\geq1$.  As a consequence spatially-compactified adjoint QCD with $N_f\geq1$ \textit{does not} have a Hagedorn instability.  In contrast, the thermal theory has a Hagedorn instability, as expected.

\subsection{Thermal compactification and the Hagedorn instability}
\label{sec:ThermalInstability}

The Hagedorn instability shows up as a singularity in the partition function at $\beta=\beta_H$, where $\beta_H$ is the first singularity encountered as $\beta$ is lowered from infinity.   The presence of the Hagedorn instability signals that the system goes through a phase transition at a temperature $T\leq T_H\equiv\beta^{-1}_H$.  This phase transition is believed to be the deconfinement transition of the gauge theory. On $S^3 \times S^1$ it was first explored in \cite{Sundborg:1999ue,Aharony:2003sx}, and was discussed in the specific context of large $N$ volume independence in \cite{Unsal:2007fb}. 

The Hagedorn singularity arises when one of the roots $r_i$ is in the unit interval $[0,1)$ and we hit a pole in \eqref{eq:ZNfThermal} as we vary $\beta$. As the circle size is decreased from $\beta = \infty$ (or $Q=0$), the first singularity occurs when $Q = r_*$, where $r_*$ is the root closest to the the origin on the unit interval. For the thermal compactification, we are guaranteed to have such a root for any $N_f\geq0$, since $P(0)=1$ and $P(1)=-4(1+N_f)$ so that there is at least one root $r_*\in [0,1)$. Furthermore the  first singularity of \eqref{eq:ZNfThermal},  $r_*$, is determined solely by the $k=1$ factor in the infinite product since for $k>1$ the singularity is at $(r_*)^{1/k}>r_*$.  The Hagedorn temperature is thus 
\begin{align}
\beta_H = - 2 R \log r_* ,
\label{eq:LHRelation}
\end{align}
and the asymptotic behavior of the thermal density of states is 
\begin{align}
\rho(E) \sim \left(\frac{1}{r_*} \right)^{E/R}\,.
\end{align}
This asymptotic behavior follows from the fact that the coefficient of a given term, say $Q^n$, in \eqref{eq:ZNfThermal} is generated by an finite product of geometric series with $k=1,\dots,n$ and is of the form
\begin{align}
\rho_n=\sum_{\{-n\leq k_{1,2,3} \leq n\}}c_{k_1,k_2,k_3}\,r_1^{k_1}r_2^{k_2}r_3^{k_3}
\end{align}
with some constants $c_{k_1,k_2,k_3}$, and the set of allowed $k_i$'s is determined by a combinatorial constraint. Then we see that asymptotically  $\rho_n\sim (1/r_*)^n$. In fact, this leading asymptotic is simply generated by the geometric series $(1-r_* Q)^{-1}$ in the infinite product \eqref{eq:ZNfThermal}, which is consistent with the statement that the Hagedorn singularity is encoded in the $k=1$ factor in \eqref{eq:ZNfThermal}. 

As explained in Appendix \ref{Ap:Roots}, the roots $r_*$ can be expressed analytically and they are given in closed form as
\begin{align}
N_f = 0: \qquad r_*  &= \sqrt{2-\sqrt{3}}\\
N_f = 1: \qquad r_*  &=  \left(\frac{1}{2}-\frac{\sqrt{2} \sqrt[4]{3}}{2}+\frac{\sqrt{3}}{2}\right) \\
N_f \geq 2: \qquad r_* &= \frac{\kappa ^2+2-\sqrt{\kappa ^4+4}}{2 \kappa },\quad\kappa\equiv \left(2\, N_f+2\, \sqrt{N_f^2-2}\right)^{1/3}\,.
\end{align}

\begin{table}
\begin{center}
  \begin{tabular}{| c || c | c | c | c | c | c | }
    \hline
    Number of flavors &  $N_f=0$  &  $N_f=1$ & $N_f=2$ & $N_f=3$ & $N_f=4$ & $N_f=5$  \\ \hline
     $R\, T_H$ &                $0.759 $ &   $0.601$ &  $0.532$  & $0.490$ & $0.461$  & $0.440$ \\ \hline
  \end{tabular}
\end{center}
\caption{Hagedorn temperatures (rounded to three digits) for the large $N$ limit of  on $S^3_R \times S^1_{\beta}$ with $N_f$ massless fermion flavors in the limit $R \Lambda \to 0$ with anti-periodic boundary conditions for fermions, so that $S^1$ is a \emph{thermal} circle. 
}
\label{table:HagedornTemperatures}
\end{table}
The corresponding Hagedorn temperatures are given in Table \ref{table:HagedornTemperatures}. Notice that with increasing $N_f$, the Hagedorn temperature decreases, as expected, since adding more degrees of freedom to the theory leads to a faster growth of density of states.  

\subsection{Spatial compactification and the disappearance of the Hagedorn instability}
\label{sec:SpatialStability}
\begin{figure*}[t]
  \centering
\includegraphics[width=\textwidth]{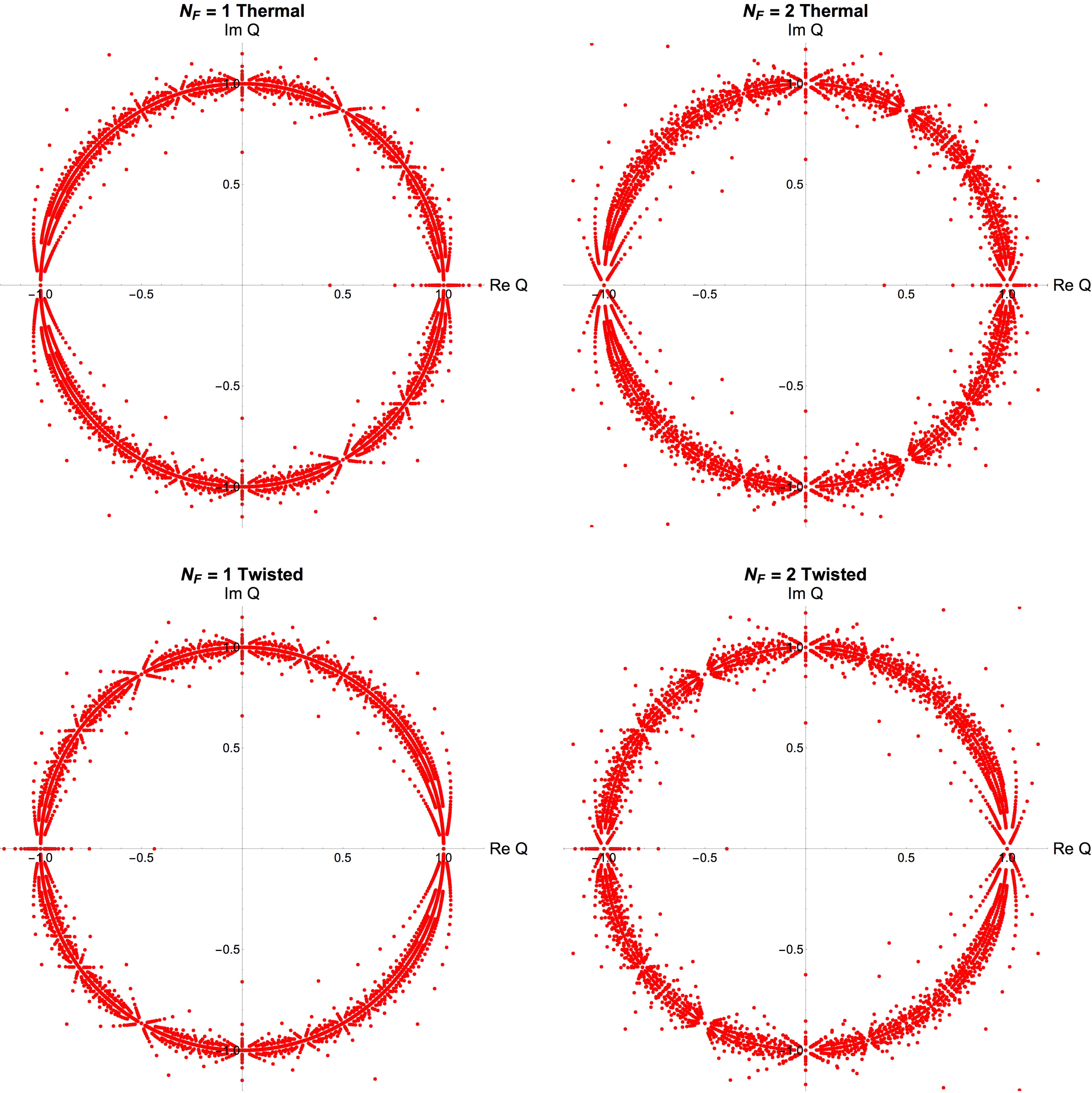}
  \caption{(Color Online.) This plot summarizes much of the paper. The red dots are singularities of the thermal (top row) and twisted (bottom row) partition functions of adjoint QCD as a function of complex temperature $Q = e^{-L/2R}$  for $N_f=1$ (left column) and $N_f=2$ (right column).  The absence of singularities on the positive real axis (except at $Q=1$, corresponding to $L=0$) is tied to the absence of Hagedorn instabilities in the twisted partition function.  The evident $Q \to -Q$ symmetry relating the singularity structure of the twisted and thermal partition follows from \eqref{eq:ZNfThermal} and \eqref{eq:ZNfTwisted}. For visual clarity we only show singularities arising from the first $30$ terms in \eqref{eq:ZNfThermal} and the first $45$ terms in \eqref{eq:ZNfTwisted}.}
  \label{fig:singularityPlot}
\end{figure*}

We now discuss the theory on a spatial circle, with periodic boundary conditions for the fermions.  The Euclidean path integral now computes the twisted partition function, $\tilde Z$, given in \eqref{eq:ZNfTwisted}.  This is the setting in which we expect large $N$ volume independence to apply\cite{Unsal:2007fb}, so the Hagedorn instability should disappear.  But  getting rid of the Hagedorn instability is hard.  It is not enough for the leading exponential behavior of the bosonic and fermionic density of states to be identical to get a twisted partition function without singularities.  There are an infinite number of  subleading exponentially-growing terms in the asymptotics of the bosonic and fermionic densities of states, and if \emph{any} of them differ there will still be a Hagedorn instability.    We now show that the degeneracies between the bosonic and fermionic states are sufficiently strong that this does not happen, and there are no Hagedorn instabilities in the twisted partition function.  The absence of instabilities as a function of $L \in \mathbb{R}^{+}$ in the twisted partition function is illustrated in Fig.~\ref{fig:singularityPlot}, which shows the locations of the poles in the twisted and thermal partition function as a function of $Q \in \mathbb{C}$.

 With a spatial $S^1$, the polynomials that appear in the denominator of $\tilde Z$ are $P\big[(-Q)^k\big]$, and the  singularities of $\tilde Z$ are determined by the roots of $\tilde P (Q)\equiv P(-Q)$,
\begin{align}
\tilde P(Q)= Q^6-3Q^4+4N_fQ^3-3Q^2+1=0\,.
\label{twisted_polynomial}
\end{align}
Given that the polynomial $Q^6-3Q^4-3Q^2+1=(Q^2+1)(Q^4-4Q^2+1)$ has only one root in $[0,1)$, and $\tilde P(0)=1$ and $\tilde P(1)=4(N_f-1)$ are both non-negative, we  see that none of roots of $\tilde P(Q)$ can be in $[0,1)$. In fact, due to the $Q\rightarrow Q^{-1}$ symmetry of \eqref{twisted_polynomial}, the only roots of $P(Q)$ along the positive real axis can be at $Q=1$. This is the case for $N_f=1$. For $N_f>1$, $P(Q)$ has 
no roots in the positive real axis at all. Furthermore, none of the factors with $k>1$ can produce singularities in $[0,1)$ either,  since those singularities are given by the $1/k^{\rm th}$ powers of roots of $\tilde P(Q)$, none of which are in $[0,1)$.  Therefore we conclude that the twisted partition function is singularity free for any $L$ and reach our main conclusion:

\begin{framed} 
Adjoint QCD on $S^3_R \times S^1_L$ with $N_f\geq1$ and periodic boundary conditions on $S^1_L$ \textit{does not} have a Hagedorn instability and stays in the confined phase for \textit{any} $L$ at $N=\infty$. 
\end{framed}

\begin{figure}[t] \centering
\includegraphics[width=0.95\textwidth]{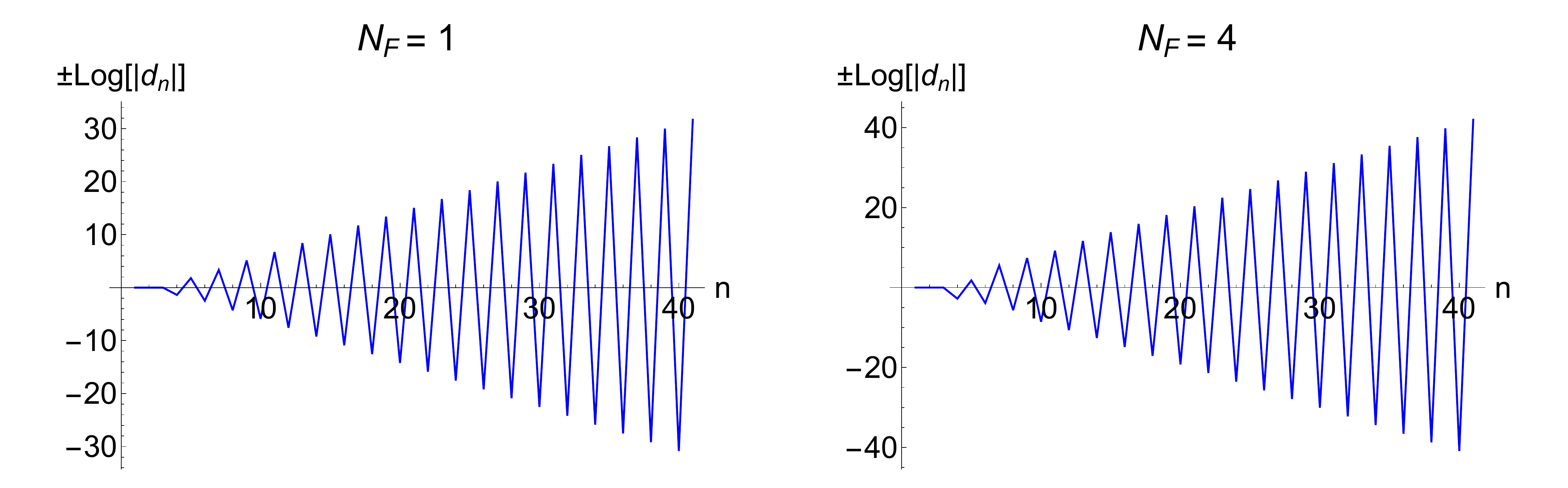} 
\caption{Logarithms of the coefficients of $Q^n$ of the series expansion of the twisted partition function $\tilde{Z}(Q)$, with $+/-$ signs for bosons/fermions.  The coefficients of even/odd powers of $Q$ are boson/fermion degeneracy factors.   We draw lines between successive data points as a visual aid to make the oscillations easier to follow.  The linearity of the envelope function means that the bosonic and fermionic densities of states both have Hagedorn growth, while the symmetry of the envelope function around zero is responsible for the elimination of Hagedorn instabilities in the twisted partition function.} \label{fig:misaligned}
\end{figure}

We now give a physical explanation for this result by taking a closer look at the the twisted and thermal partition functions. The coefficients of $Q^n$ in $\tilde{Z}$ count the number of bosonic states minus the number of fermionic states at energy $E_n = n/(2R)$, while in $Z$ they count the number of bosonic states plus fermion states. The states counted by even powers of $Q$ are purely bosonic, while states counted by odd powers of $Q$ are purely fermionic.\footnote{The same result also follows from the fact that in the $R \Lambda \to 0$ limit, the energy of a given bosonic/fermionic state is simply given by the radial quantum number of the vector/spinor $S^3$ spherical harmonic function, i.e.
\begin{align}
\omega_{B,n}=  \frac{n+1}{R},\qquad  
\omega_{F,n} = \frac{n+\frac{1}{2}}{R} \nonumber
\end{align}
Since $Q^n=e^{-2 L \omega_n}$,  even/odd powers of $Q^n$ correspond to bosonic/fermionic states respectively.  }  
Expanding the partition functions in $Q$ with e.g. $N_f=1$ yields
\begin{align}
\tilde{Z}_{N_f=1}(Q) &= 1-4 Q^3+6 Q^4-12 Q^5+28 Q^6-72 Q^7+168 Q^8-364 Q^9+828 Q^{10} + \cdots\\
Z_{N_f=1}(Q) &= 1+4 Q^3+6 Q^4 + 12 Q^5+28 Q^6 + 72 Q^7+168 Q^8 + 364 Q^9+828 Q^{10} + \cdots
\label{eq:ZTwistNf1Series}
\end{align}
The coefficients $\rho_n$ of $Q^n$ grow rapidly with $n$ and reach their asymptotic behavior $\rho_n\sim (1/r_*)^n$ quickly.

\begin{figure}[t] \centering
\includegraphics[width=0.8\textwidth]{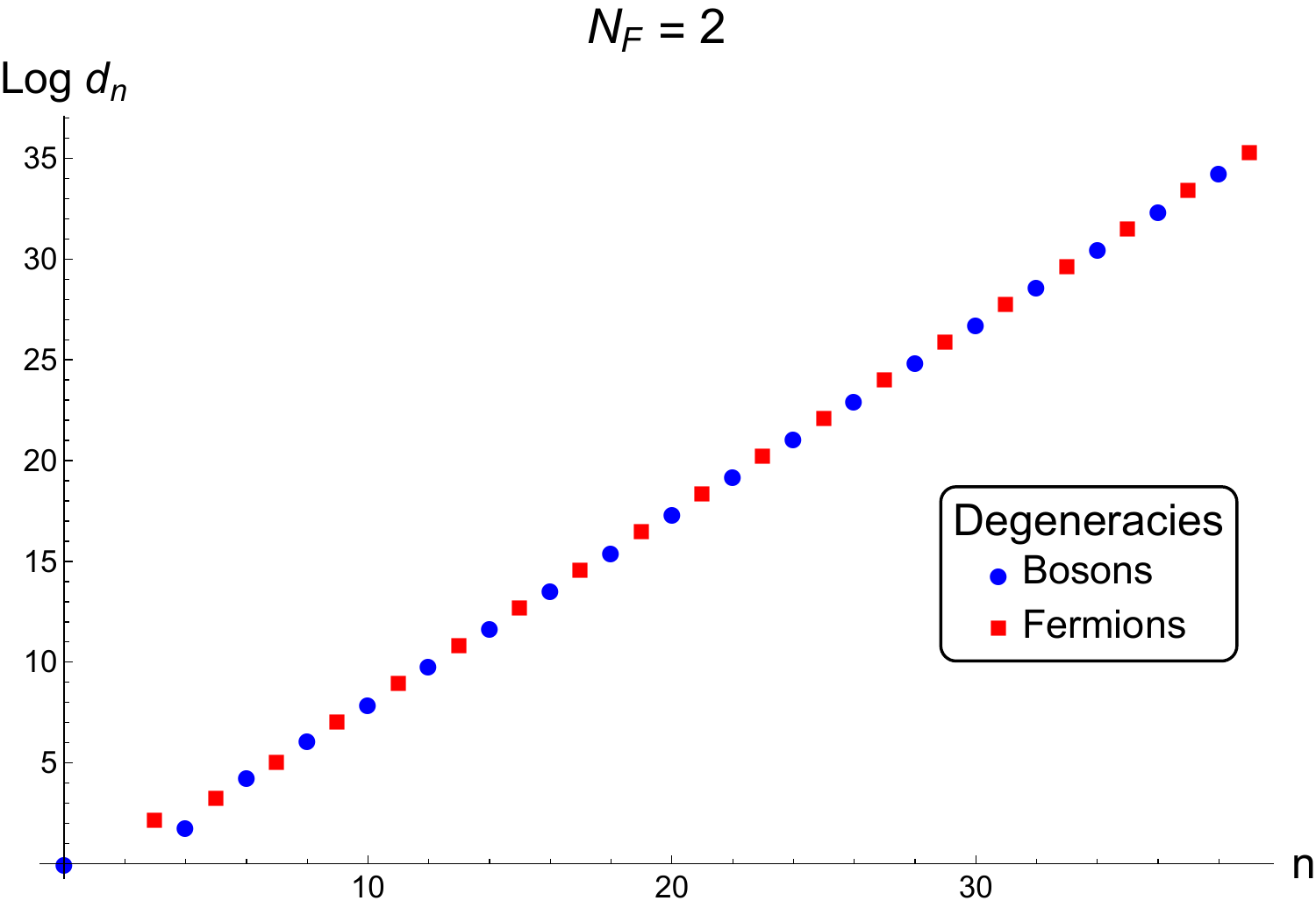} 
\caption{Logarithms of the coefficients of $Q^n$ of the series expansion of the thermal partition function $Z(Q)$ for $N_F=2$.  The bosonic and fermionic state degeneracy factors have identical asymptotic scaling with $n$. } \label{fig:BFcoefficients}
\end{figure}

As illustrated in Fig.~\ref{fig:BFcoefficients},  where we plot the logarithms of $d_n$ for $N_f=2$, the asymptotic behavior of bosonic and fermionic density of states is identical. The sole difference between the thermal and the twisted case is that 
\begin{align}
d^{\rm twisted}_n = (-1)^n d^{\rm thermal}_n
\label{eq:TwistedFromThermal}
\end{align}
where $d^{\rm twisted/thermal}_n$ are the coefficients of $Q^n$.  This is of course an obvious consequence of the definitions.  What is far less obvious a priori is that as illustrated in Fig.~\ref{fig:BFcoefficients}, it appears that both the bosonic and fermionic degeneracy factors in the thermal partition function  can be thought as coming from the \emph{same} smooth function of $n$, which becomes monotonic past some $n = n_*$ (in the figure $n_* = 4$).   This apparent underlying function gets sampled at even integers to give the bosonic degeneracies, and gets sampled at the odd integers to give the fermionic degeneracies.     If an analytic continuation of $d_n$ to a function $f(n)$ of $n \in \mathbb{C}$ were to be found explicitly and could be shown to be monotonic, it would be one way to  demonstrate that the bosonic and fermionic hadronic states are entirely degenerate up to an offset due to the curvature for any $N_f$.   We leave this challenging task to future work, since in our view understanding the degeneracy pattern in terms of symmetries may be more directly illuminating.   

From Fig.~\ref{fig:misaligned} and Fig.~\ref{fig:BFcoefficients} it is clear that the  $d^{\rm twisted}_n$ coefficients form an alternating sequence with a symmetric envelope around zero.  These oscillations, illustrated in Fig.~\ref{fig:misaligned}, are behind the disappearance of the Hagedorn instability for the spatial compactification. 

We note that this type of cancellation mechanism of bosonic and fermionic contributions to the twisted partition function is rather different than the more familiar ``supersymmetry-like'' fermion-boson cancellations, which occur \emph{within} each given energy level. The cancellations we see in adjoint QCD on $S^3 \times S^1$ instead involve repeated cancellations neighboring levels of bosons and fermions.  The same effect was seen in work on misaligned supersymmetry \cite{Kutasov:1990sv,Dienes:1994np,Dienes:1994jt,Dienes:1995pm}, and we discuss the connection between adjoint QCD and misaligned supersymmetry in Section~\ref{sec:Misaligned}. Note however that the offset between the bosonic and fermionic degeneracies which leads to the oscillations is due to the $S^3$ curvature.  If $R\Lambda \gtrsim 1$ the curvature should become unimportant, and the boson-fermion cancellations should start taking place within each level if the theory still lacks a Hagedorn instability, as discussed in \cite{Basar:2013sza}.

\subsection{Twisted Casimir energy in adjoint QCD}
\label{sec:Casimir}
In this section we compute the twisted vacuum energy 
\begin{align}
\tilde{C} = C_B - C_F
\end{align}
where $C_{B}, C_{F}$ are the vacuum energies due to the bosonic states and $C_F$, which can be computed from the behavior of the twisted partition function.  Since we are working on $S^3 \times S^1$, these vacuum energies can be thought of as Casimir energies on $S^3$, motivating the notation.    The computation of Casimir energies $C = C_B + C_F$ in large $N$ gauge theories on $S^3 \times S^1$ with thermal boundary conditions involves similar techniques but is more involved, and is discussed in a separate paper\cite{Basar:2014hda}.

To begin, recall that the physical states of this large $N$ theory are single-trace operators, and their energies and degeneracies are counted by the twisted single-trace partition function from \eqref{eq:ZST}
\begin{align}
\label{eq:ZSTrepeated}
\tilde{Z}_{\rm ST}[q]  &= -\sum_{m=1}^{\infty} \frac{\varphi(m)}{m} \log \left[1-z_V(q^{m}) +N_f z_F(q^{m})\right] \\
& \equiv \sum^{\infty}_{n=1} D_n e^{-L \omega_n}
\end{align}
and $\omega_n = n/(2R)$ is the energy of the $n$-th mode with degeneracy $D_n$.  Let us define
\begin{align}
\tilde{C}(L) \equiv -\frac{1}{2}{\partial \tilde{Z}_{\rm ST}\over \partial L}&=  \frac{1}{2}\sum^{\infty}_{n=1}D_n \omega_n e^{-L \omega_n}\,.
\label{eq:Cbeta}
\end{align}
Then the twisted Casimir energy\footnote{We emphasize that this definition relies on using the $N$ independent spectrum obtained after large $N$ limit being taken first. We thank O.~Aharony, C.~P.~Herzog, and M.~Yamazaki for discussions on this point.} can be formally written as
\begin{align}
\tilde{C}=  \frac{1}{2}\sum^{\infty}_{n=1}D_n \omega_n= -\frac{1}{2}{\partial \tilde{Z}_{\rm ST} \over \partial L} \big|_{L=0}=\tilde{C}(0)\,.
\label{eq:CasimirFormula}
\end{align}
Of course this formal expression is divergent and has to be regularized and renormalized to extract the physical quantity $\tilde{C}$. Thanks to the absence of any phase transitions as $L$ is varied, $\tilde{L}(C)$ is well-defined for any $L \neq 0$, and can be viewed as defining as a spectral regularization of the divergent sum in $\tilde{C}$.   The structure of the singularities in the twisted single-trace partition function is illustrated in Fig.~\ref{fig:singularityPlot} for $N_f=1$ and $N_f=2$.  The absence of any singularities on the positive real axis makes it easy to take the $L \to 0$ limit above. The situation is more subtle for thermal compactifications, see \cite{Basar:2014hda} for a full discussion.

Our renormalization prescription amounts to isolating the divergent part of $\tilde{C}(L\rightarrow0)$ and extracting the $L$ independent, finite part.  The divergent part of $\tilde{C}(L)$, which scales with the UV cutoff $\mu$ as $\mu^2/R^2$~\footnote{The absence of a $\mu^4$ divergence is itself quite interesting.  See \cite{DiPietro:2014bca} for a related recent discussion in the context of supersymmetric QFTs.} is absorbed by a $\mu^2 \int d^4 x \sqrt{g} \, \mathcal{R}$ counter-term, and since the only divergence is a power law there are no issues with cutoff scheme dependence.

We now evaluate the twisted Casimir energy in two different ways. First, we use a hybrid zeta function and heat-kernel-like regularization procedure to extract the finite part of $\tilde{C}(L\rightarrow0)$ analytically. Second, we directly evaluate $\tilde{C}(L)$ numerically, and confirm the findings of the analytical manipulations. The details of the numerical computation are explained  in Appendix \ref{Ap:Casimir}. In both cases we find that the finite, $L$-independent part of $\tilde{C}( L\rightarrow0)$ vanishes and conclude that the twisted Casimir energy of adjoint QCD on $S_R^3\times S_L^1$ at $N=\infty$ and small $R$ is \emph{zero} for any  $N_f\geq1$.

To compute $\tilde{C}$ we need to understand the $L\rightarrow0$ limit in \eqref{eq:CasimirFormula}, and to this end we first isolate the part of the sum from \eqref{eq:ZSTrepeated} in $\partial \tilde{Z}/\partial{L}$ which is divergent:
\begin{align}
{1\over 4 R} Q{ \partial \over \partial Q}  \log \left[1-z_V(Q^{2m}) +N_f z_F(Q^{2m})\right] = &{1 \over 2R}\left(\frac{3 m Q^{2 m} \left(2 N_f Q^m-2 Q^{2 m}+Q^{4 m}-1\right)}{Q^{2 m} \left(4 N_f Q^m-3 Q^{2 m}+Q^{4 m}-3\right)+1}\right)\nonumber \\ &+{3 \over 2R} \frac{ m Q^{2 m}}{1-Q^{2 m}}
\label{eq:separation}
\end{align}
We can take $Q=1$ in the first term since the divergent part is isolated in the second term.\footnote{For $N_f=1$, the separation of the divergent and finite part in \eqref{eq:separation} is different. However the $N_f$ dependence drops out in the final answer for the twisted Casimir energy for arbitrary $N_f$. So, taking $N_f=1$ at the end of the calculation, as presented above, is safe.}  Doing so, we arrive at the expression
\begin{align}
\tilde{C}(L\rightarrow0)= -{3\over 4 R}\sum_{m=1}^{\infty} \varphi(m) -{3 \over 2R} \lim_{L\rightarrow0}\sum_{m=1}^{\infty} \varphi(m)  \frac{Q^{2 m}}{1-Q^{2 m}}\,.
\end{align}
Both of these expressions are formally divergent. Regulating the first term using the zeta-function identity $\sum_{m=1}^\infty \varphi(m) m^{-s}=\zeta(s-1)/\zeta(s)$, and using a Lambert series identity $\sum_{m=1}^\infty\varphi(m) q^m/(1-q^m)=q/(1-q)^2$ for the second term, leads to the result
\begin{align}
\tilde{C}(L\rightarrow0)&= -{3\over 4 R}{\zeta(-1)\over\zeta(0)}-{3 \over 2R} \lim_{L\rightarrow0} \frac{Q^{2}}{(1-Q^{2})^2}  \nonumber\\
&=-{3\, \zeta(-1)\over 4\,\zeta(0)R}+{1\over 8R}-{3 R \over2 L^2}=-{3 R \over2 L^2} + 0 \times L
\label{eq:CasimirResult}
\end{align}
The fact that the $L$-independent term vanishes yield the conclusion that the twisted Casimir energy vanishes. 

Two remarks about the calculation above are in order. First, in principle, one might be worried about the algebraic manipulations such as splitting terms in formally divergent sums and regularizing them individually. This is not an issue because $\tilde{C}(L)$ is finite for any finite $L$.   Moreover, even if \eqref{eq:separation} is not viewed in the context of being embedded in the regularized expression $\tilde{C}(L)$, note that both of the regularizations leading to \eqref{eq:CasimirResult} involve cutoff functions which only depend on the energy spectrum, justifying the manipulations. Second, one might be concerned that the  $L^{-2}$ terms in the analytical calculation above and in the numerical computation in Appendix \ref{Ap:Casimir} are different. This is not issue, because only the finite $L$-independent terms are physical and regulator independent. The divergent pieces do not have to agree if different regulators are used.  The numerical calculation extracts $\tilde{C}$ directly from the scaling of $\tilde{C}(L)$ at small $L$, while the analytic calculation brings in a zeta function along the way, which amounts to a modification of the regularization scheme and a corresponding difference in the coefficients of the divergent pieces in the two computations.  
 
The underlying physical reason for the remarkable result that the twisted Casimir energy is zero is not known to us, but presumably it is a consequence of previously unrecognized symmetries of large $N$ adjoint QCD, as are the rest of our results. We note that it is actually expected from the fact that the twisted partition function of \eqref{eq:FullPartitionSpatial} has a $T$-reflection symmetry with a zero vacuum energy, as noted in \cite{Basar:2014mha}.  A more detailed exploration of the very interesting interplay between $T$-reflection symmetry and the vacuum energy of confining large $N$ gauge theories on $S^3 \times S^1$ is discussed in \cite{Basar:2014hda}. 

\subsection{Relation to  misaligned supersymmetry}
\label{sec:Misaligned}
We have seen that the way spatially compactified adjoint QCD on $S^3 \times S^1$ escapes the Hagedorn instability involves cancellations between the bosonic and fermionic densities of states, both of which grow exponentially, and the cancellations arise due to an oscillation between the number of bosonic and fermionic states at successive excitation levels.    

These cancellations fit the framework of `misaligned supersymmetry' developed in \cite{Dienes:1994np,Dienes:1994jt,Dienes:1995pm}.  These papers explored the structure of the partition functions of perturbative fundamental closed string theories.  Consistent closed string theories are always modular-invariant, but may or may not have spacetime supersymmetry.   Refs.~\cite{Dienes:1994np,Dienes:1994jt,Dienes:1995pm} pointed out that modular invariance along with the absence of tachyons implies certain intricate patterns of relations between the degeneracies of bosonic and fermionic states.  These relations imply that the leading exponentially-growing parts of the bosonic and fermionic densities of states in the closed string theories cancel against each other in the twisted partition function.  With spacetime supersymmetry, the cancellations occur within each level.  More generally, however, for modular-invariant string partition functions without spacetime supersymmetry, these cancellations are due to sign-oscillating mismatches between bosonic and fermionic state degeneracies\cite{Dienes:1994np,Dienes:1994jt,Dienes:1995pm}.  Misaligned supersymmmetry can also imply the vanishing of super-traces which contribute to the cosmological constant and its divergences\cite{Dienes:2001se}.   

Such oscillating cancellations between bosonic and fermionic states are exactly what we have seen in our analysis.  In this sense, large $N$ adjoint QCD on $S^3 \times S^1$ with $N_f \ge 1$ appears to give the first known field-theoretic realization of the string-theoretic idea of misaligned supersymmetry.   This raises many interesting questions.  For instance, in the analysis of \cite{Dienes:1994np,Dienes:1994jt,Dienes:1995pm} the modular invariance of the partition functions of string theories played a starring role.  Large $N$ gauge theories are believed to be describable as some kind of weakly-coupled string theories, so if adjoint QCD enjoys a realization of misaligned supersymmetry, one might wonder whether its partition function enjoys some form of modular invariance.  If the partition function were to be modular invariant, the would yield an underlying reason for the pattern of cancellations.  We now explore this possibility.

\begin{figure}[tbp] \centering
\includegraphics[width=0.85\textwidth]{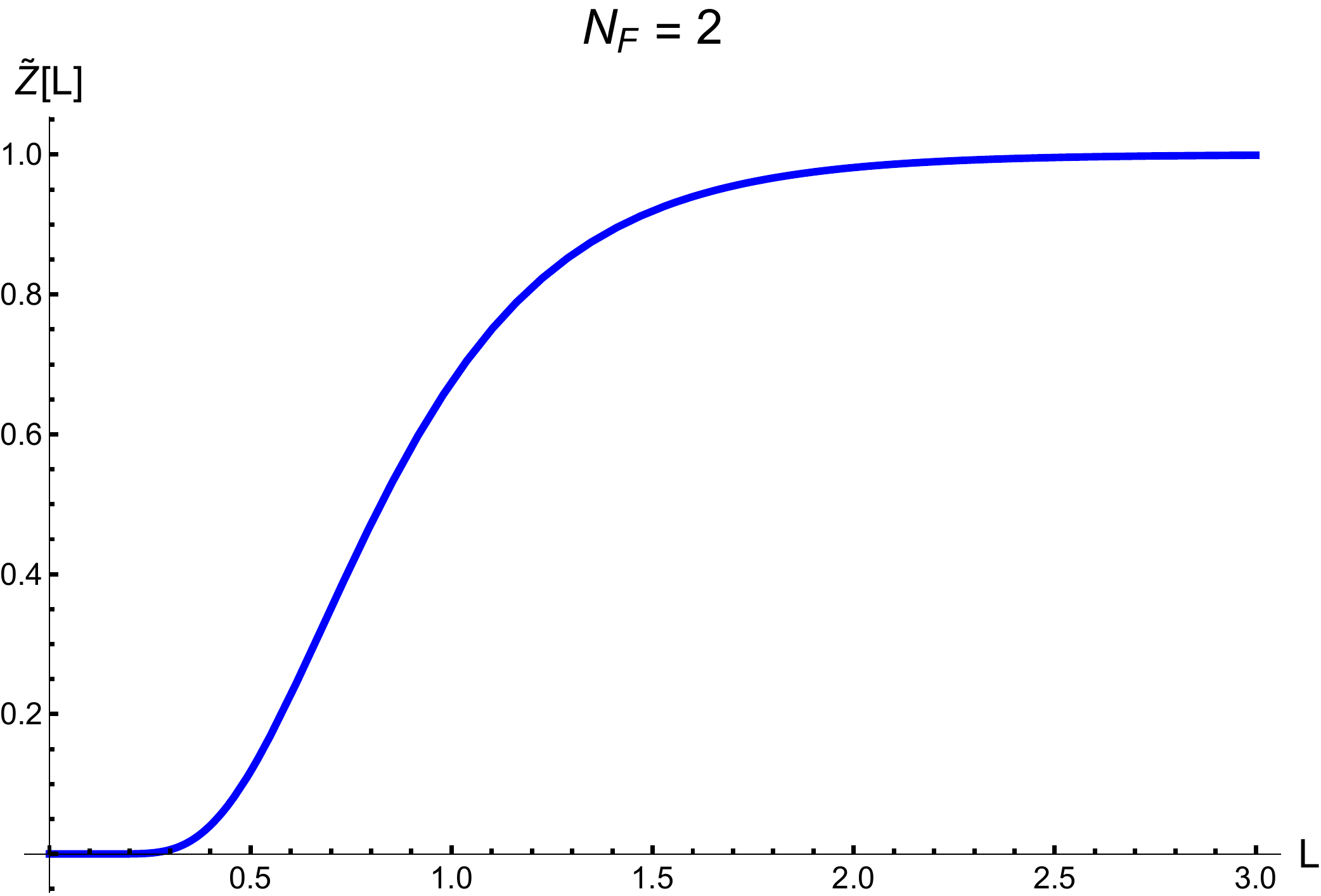} 
\caption{Plot of the $N_f=2$   partition function on $S^3_R \times S^1_L$ when $R \Lambda \ll 1$, which illustrates the lack of invariance under $L\rightarrow {c\over L}$ for any $c >0$.  The fact that the confined-phase twisted partition function is well-defined and continuous for any $L \sim N^0$ is a consequence of massive cancellations between bosons and fermions. }
\label{fig:Nf1Z}
\end{figure}

Modular invariance of a partition function $Z$ for a theory on a spatial circle implies
\begin{align}
Z(\tau) = Z(\tau +1)  = Z(-1/\tau)
\end{align}
where $\tau$ is defined through $Q = e^{2\pi i \tau} = e^{-{L\over 2 R}}$.   Hence modular invariance implies $Z(L) = Z[(4\pi R)^2/L]$, which is a manifestation of $T$-duality.    However, the twisted partition function \eqref{eq:ZNfTwisted} does not have modular invariance. The simplest way to see this is to observe that $\tilde Z(L)$ does not have the right shape for modular invariance, as is illustrated in Fig.~\ref{fig:Nf1Z}, since it has different limits for $L \to 0$ and $L\to \infty$, approaching $0$ and $1$ respectively.  We can also see the lack of modular invariance algebraically.  By using the modular properties of the Dedekind function, and Jacobi's transformation identities for the theta functions, it can be shown that under the two generators of $SL(2,Z)$ modular transformations
\begin{align}
T : \; \tau \to \tau +1 \qquad S:\; \tau \to -1/\tau
\end{align}
where $\tau$ is assumed to be in the upper half-plane, the full partition function transforms as
\begin{align}
\tilde{Z}_{\rm QCD[Adj]} (\tau +1) &= \tilde{Z}_{\rm QCD[Adj]} (\tau) \\
\tilde{Z}_{\rm QCD[Adj]} (-1/\tau) &=  \left(-i \tau\right)^{3/2}  {\eta^3(\tau/2)\over \eta^3(2\tau)} \left(\prod_{i=1}^{3} \frac{e^{i \tau \nu_i^2/\pi}\vartheta_2(\nu_i|e^{i\pi\tau})}{\vartheta_4(\tau \nu_i | e^{i\pi\tau})}\right) \tilde{Z}_{\rm QCD[Adj]} (\tau)
\end{align}
where $e^{2i\nu_i}=r_i$. This means that the partition function of large $N$ adjoint QCD on $S^3 \times S^1$ is \emph{not} invariant under the $SL(2,\mathbb{Z})$ modular group, nor does it transform as a modular form.  However, as discussed extensively in e.g. \cite{Dienes:1994np}, closed string partition functions are made from special combinations of both holomorphic and antiholomorphic (in $\tau$) modular functions.  As a result the modular invariance of closed string theories is intimately related to the fact that string partition functions include contributions from `off-shell' states with $m \neq n$ where $(m,n)$ are the world-sheet energies of (left, right) moving states.  Such states do not appear in field theory, so one should not normally expect that modular invariance would show up in any simple way in a field theory partition function, even if the field theory has a dual description as a string theory with modular invariance.\footnote{We thank K. Dienes for explaining this to us.}  Nevertheless, it would be interesting to explore whether our results are some sort of field-theoretic remnant of misaligned supersymmetry in the string dual of adjoint QCD.

 Of course, we are dealing with a weakly-coupled limit of adjoint QCD, so the phenomena we are seeing should have a description directly within field theory in any case.   While it would be wonderful to understand the string theory dual of the adjoint QCD, there should be no need to do this to understand the pattern of degeneracies between bosonic and fermionic states that we have seen.   In the next section we make some remarks on how our results may be understood directly in field theory through emergent fermionic symmetries.

\section{Emergent fermionic symmetries in adjoint QCD on $S^3 \times S^1$}
\label{sec:Symmetries}
In this section we comment on the relation between our results and the notion of emergent fermionic symmetries in the large $N$ limit.  Understanding these relations is especially important for seeing whether our results will continue to hold once we move away from the $R \Lambda \to 0$ limit, where $\lambda \to 0$. 

\subsection{$N_f=1$}
\label{sec:Nf1Symmetry}
$SU(N)$ massless adjoint QCD in flat space with $N_f=1$ has $\mathcal{N}=1$ supersymmetry, since it is just $\mathcal{N}=1$ super-Yang-Mills theory.   However, the supersymmetry is broken on $S^3 \times S^1$ due to the curvature couplings.  On a curved generic manifold there are no covariantly constant spinors, so there is no way to define conserved supercharges.   
The exception is when the compactification manifold has enough isometries \emph{and} the field theory has a non-anomalous continuous $\mathcal{R}$ symmetry.\footnote{Then one can define a `twisted' subgroup of the Lorentz symmetry which lives in a diagonal subgroup of isometry transformations and $\mathcal{R}$ symmetry rotations, and at least some fraction of the original supersymmetry can be preserved in the compactified theory.  For discussions of how this works for theories with $\mathcal{N} \ge 1$ supersymmetry on $S^3 \times \mathbb{R}$ and $S^3_R \times S^1_L$ see \cite{Sen:1985ph,Romelsberger:2005eg,Festuccia:2011ws,Dumitrescu:2012ha}.}
In general, 4D $\mathcal{N}=1$ SUSY QFTs have a classical $U(1)_{\mathcal{R}}$ global symmetry.  When a 4D $\mathcal{N}=1$ theory is compactified on $S^3_{R} \times \mathbb{R}$, the SUSY algebra is modified from its flat-space form to (see e.g. \cite{Sen:1985ph}):
\begin{align}
\{Q_{\a}, \bar{Q}_{\da} \} &= 2 i \sigma^{\mu}_{\a \da} \partial_{\mu} - \frac{2}{R} \sigma^{0} n_{\mathcal{R}}
\label{eq:TwistedSUSY}
\end{align}
Here $n_{\mathcal{R}} = \int d^{3} x\, j^{0}_{\mathcal{R}}$ is the charge operator associated with the $U(1)$ $\mathcal{R}$-current $j^{\mu}_{\mathcal{R}}$. Under the $\mathcal{R}$ symmetry, gauge fields have charge zero, while the Weyl fermions have charge $1$.  Hence when there is an unbroken continuous $\mathcal{R}$ symmetry in the full quantum theory, supersymmetry is preserved on $S^3 \times \mathbb{R}$ and on $S^3 \times S^1$ with periodic boundary conditions.  

This setup does not work for $\mathcal{N}=1$ $SU(N)$ SYM, since it suffers from a chiral anomaly that breaks $U(1)_{\mathcal{R}} \to Z_{2N}$.  So there is no continuous $\mathcal{R}$ symmetry.\footnote{On $\mathbb{R}^4$, there is a further spontaneous breaking of the non-anomalous part of the $\mathcal{R}$-symmetry down to $\mathbb{Z}_2$.} As a result the classical supersymmetry  of $N_f=1$ $SU(N)$ adjoint QCD on $S^3_R \times \mathbb{R}$ or $S^3_R \times S^1_L$ suffers from an anomaly, and the theory has no fermionic symmetries except in the $\mathbb{R}^4$ limit. 

This raises a puzzle, because $N_f=1$ adjoint QCD on $S^3_R \times S^1_L$ with $R\Lambda \ll 1$ has unbroken center symmetry for any $L \sim N^0$, enjoys large $N$ volume independence, and has no Hagedorn instabilities for any $L$.  The absence of Hagedorn instabilities is due to conspiracies between the bosonic and fermionic densities of states which amount to relations between degeneracies and energies of an infinite number of bosonic and fermionic states. As argued in the introduction and in \cite{Basar:2013sza}, this seems to call for a symmetry.  And yet we have just said that the $SU(N)$ $N_f=1$ theory definitely has no fermionic symmetries.  What is going on?  We now argue that the resolution of the puzzle is that there is an emergent large $N$ fermionic symmetry.

Recall that the chiral anomaly for the would-be conserved current $j_{\mu}^{\mathcal{R}}$ is
\begin{align}
\partial^{\mu} j^R_{\mu} = \frac{\lambda}{16\pi^2} \Tr F_{\mu \nu} \tilde{F}^{\mu \nu} ,
\label{eq:anomaly}
\end{align}
This anomaly equation has no manifest $1/N$ suppression factors, and $U(1)_{\mathcal{R}}$ breaking appears to be unsuppressed at large $N$. While this is true, there are some important subtleties on $S^3_R \times S^1_L$ with $R \Lambda \ll 1$, the regime in which we are working.   Note that these subtleties can be argued to be negligible strictly at $R \Lambda \to 0$, but become important as soon as we allow $\lambda$ to be finite.

It is useful to recall the reason for the anomaly breaking pattern $U(1)_{\mathcal{R}} \to \mathbb{Z}_{2N}$.  The origin of the unbroken $\mathbb{Z}_{2N}$ factor lies in the fact that the right-hand side of the anomaly equation is a total derivative, and is only non-zero on instanton field configurations with non-zero topological charge $Q$\footnote{In the 't Hooft large $N$ limit, at $N=\infty$ there is no spontaneous breaking of the gauge group to the Cartan subgroup, even in a confining background, so we do not expect well-defined monopole-instanton field configurations with fractional topological charge to appear and interfere with our argument.}.  But in the $N_f=1$ theory the instantons  carry  $2N |Q|$ fermion zero modes, and generate effective  't Hooft vertex interactions for the fermions which break $U(1)_{\mathcal{R}}$ but are invariant under its $\mathbb{Z}_{2N}$ subgroup.  So the interacting theory only enjoys the $\mathbb{Z}_{2N}$ symmetry.   On the one hand, at large $N$, a $\mathbb{Z}_{2N}$ symmetry ought to have the same power as a $U(1)$ symmetry, up to $1/N$ corrections.  This makes it appear that the anomaly is suppressed at large $N$.  On the other hand, the anomaly cannot be suppressed, because the RHS of Eq.~\eqref{eq:anomaly} is unsuppressed relative to the LHS.  

Despite first appearances, these observations are not in conflict with each other.  To get a non-vanishing contribution from the right-hand side of Eq.~\eqref{eq:anomaly} one must consider correlation functions with enough fermion operators to saturate the $2N |Q|$ zero modes. Let us call color-singlet operators with $\gtrsim N^1$ fermionic operators inside the color trace `heavy', and call operators which have $\sim N^0$ fermions `light'.  The fact that this distinction can be made relies on the fact that  in the regime we are considering, $R\Lambda \ll 1$, there is no chiral condensate, so there is no spontaneous breaking $\mathbb{Z}_{2N} \to \mathbb{Z}_2$. So it makes sense to classify operators by their $\mathbb{Z}_{2N}$ charge when $R \Lambda \ll 1$.\footnote{To see this recall that the fermions have a effective curvature-induced $\mathcal{R}$-symmetry-preserving mass $1/(2R)$.   This implies that e.g. the two-point correlation function falls off exponentially:
\begin{align}
\langle \lambda \lambda(t) \; \lambda \lambda(0) \rangle \sim e^{-t/(2R)}
\end{align} 
So there is no long-range order, meaning that there is no spontaneous breaking of the discrete remnant of the $\mathcal{R}$ symmetry.  Note as well that the absence of spontaneous symmetry breaking is not happening for trivial Coleman-Mermin-Wagner reasons, since we are working at large $N$. }      It does not make sense to do so if $R \Lambda \gtrsim 1$, because then the $\mathbb{Z}_{2N}$ symmetry becomes spontaneously broken due to the formation of a gluino condensate. 

The $N$-independence of the right-hand side of Eq.~\eqref{eq:anomaly}  means that for heavy states the $\mathcal{R}$-symmetry is irredeemably broken.  There is no reason to expect their energies and degeneracies to be related to each other by any fermionic symmetry.  But consider states  whose interpolating operators are light.  Correlators of light operators cannot saturate the instanton zero modes, so for these states the $\mathbb{Z}_{2N}$ symmetry gives non-trivial relations.    At large $N$, as far as these light states are concerned,  the theory enjoys a $U(1)_{\mathcal{R}}$ symmetry.    These light states are precisely the ones that are important throughout our analysis of partition functions with $L \sim N^0$.\footnote{If $L \sim N^{-1}$, states with energies of order $N$ start to participate in the partition function, and volume independence is expected to be lost on very general grounds.  This fits nicely with our discussion here:  the emergent symmetry should stop being effective once $L$ becomes of order $1/N$.}   So \emph{when acting on states that remain light at large $N$}, the SUSY algebra in Eq.~\eqref{eq:TwistedSUSY} is anomaly-free up to $1/N$ corrections. 

The punchline should now be clear:  $N_f=1$ adjoint QCD on $S^3 \times S^1$ has an emergent fermionic symmetry in the large $N$ limit, even when $\lambda$ is finite and $R \Lambda$ 
is not sent to zero, so long as $R\Lambda \lesssim 1$ and there is no gluino condensate. 
 Not coincidentally, we expect that it also enjoys large $N$ volume independence, with no Hagedorn instabilities in the twisted partition function thanks to massive cancellations between bosonic and fermionic densities of states, even when $R \Lambda$ is finite.  
 We have explicitly verified this expectation in limit $R \Lambda \to 0$ in the preceding sections.   

When $R\Lambda \gg 1$, we expect a gluino condensate to form, which invalidates our arguments for an emergent supersymmetry.  Hence one expects an explicit breaking of the supersymmetry.  However, this breaking should be suppressed by powers of $e^{-R\Lambda}$, and supersymmetry will be restored in the flat-space limit.  
When $R\Lambda \lesssim 1$, we just argued that at $N=\infty$ supersymmetry will be an emergent symmetry.  What is left unclear is what happens when $R \Lambda \sim 1$ at $N=\infty$.  In this regime  one would expect that $\mathbb{Z}_{2N}$ will break to $\mathbb{Z}_2$ due to the formation of a gluino condensate, and the arguments we gave above no longer apply.   
Whether this can be accompanied by a breakdown of the sort of cancellations we have seen in our analysis is beyond the scope of this paper.

\subsection{$N_f>1$}
The relations we saw between the spectrum of bosonic and fermionic hadronic excitations in adjoint QCD on $S^3 \times S^1$ are very similar for $N_f=1$ and $N_f>1$.  Here we comment on the symmetries of adjoint QCD for $N_f>1$.  First, note that at the microscopic level, adjoint QCD has $2(N^2-1)$ bosonic degrees of freedom (from the gluons) and $2 N_f (N^2-1)$ fermionic ones (from the quarks).  Once $N_f >1$, something more exotic than the story in Sec.~\ref{sec:Nf1Symmetry} is necessary due to the mismatch in the number of microscopic degrees of freedom. It seems that any emergent fermionic symmetry could  not be a standard supersymmetry.  What could it look like?\footnote{We are very grateful to D.~Dorigoni for collaboration on the material in this section at an early stage.}

At the moment we can only make a suggestive observation in this direction. In the preceding sections we saw that the $\lambda \to 0$ limit of adjoint QCD on $S^3 \times S^1$ is already very interesting, with many of the features of the $\lambda>0$ theory (such as confinement) remaining qualitatively preserved.  With this as an inspiration we examine the $\lambda = 0$ limit of adjoint QCD in \emph{flat space} and show that it has a fermionic symmetry for any $N_f \ge 1$.   The Lagrangian density of the theory is
\begin{align}
\mathcal{L} = \frac{1}{g^2}\Tr \left[ -\frac{1}{2} F^2 + 2 i \sum_{a = 1}^{N_f} \left( \, \bpsi_{a \da} \bsigma^{\mu \da \a} D_{\mu} \psi_{a \a}  \right) \right]
\label{eq:QCDAdjAction}
\end{align}
where $\psi_{a \a}, a = 1, \ldots, N_f$, $\alpha$ is a spinor index is an adjoint Weyl fermion, and  $F_{\mu \nu} = \partial_{\mu} A_{\nu} - \partial_{\nu} A_{\mu} - i[A_{\mu},A_{\nu}], D_{\mu} \psi^a = \partial_{\mu} \psi^a - i[A_{\mu},\psi^a]$.  The equations of motion are
\begin{align}
D_{\mu} F^{\mu \nu} &= [\bpsi^a_{\da} \bsigma^{\nu \da \a}, \psi_{a, \a}],  \qquad
\bsigma^{\mu \da \a }D_{\mu} \psi_{a \a} = 0 .
\label{eq:EoM}
\end{align}
We now exhibit field variations that lead to a fermionic symmetry in the $\lambda = 0$ limit for any $N_f \ge 1$.  The variations are proportional to $N_f$ infinitesimal Weyl fermion parameters $\bepsilon^a, \epsilon_a$:
\begin{align}
\delta A^{\mu}(x) &= -\frac{1}{\sqrt{2}} \left[ \bepsilon^{a}_{\da}  \bsigma^{\mu \da \a} \psi_{a \a}(x) +  \bpsi^a_{\da}(x)  \bsigma_{\mu}^{\da \a} \epsilon_{a \a} \right] \\
\delta \psi_{a \a}(x) &= \frac{-i}{2\sqrt{2} } \sigma^{\mu}_{\a \db} \bsigma^{\nu \db \b}  \epsilon_{a \b}  \,F_{\mu \nu} (x)  \\
\delta \bpsi^{a}_{\da}(x) &= \frac{+i}{2\sqrt{2} } \bepsilon^{a}_{\db} \bsigma^{\nu \db \a} \sigma^{\mu}_{\a \da} \,  F_{\mu \nu}(x) 
\label{eq:superSUSYvariations}
\end{align}
Note that for $N_f=1$ these are simply the $\lambda = 0$ limit of the standard on-shell $\mathcal{N}=1$ SUSY transformations.  To check the variation of the action, we write
\begin{align}
\delta \mathcal{L} = \delta \mathcal{L} |_{\rm gauge}  +  \delta \mathcal{L} |_{\rm fermion}
\end{align}
where
\begin{align}
\delta \mathcal{L} |_{\rm gauge} &= \frac{-1}{2} \Tr \left[ 2 F_{\mu \nu}  \delta\left( F^{\mu \nu}  \right)\right] 
\end{align}
and
\begin{align}
\delta \mathcal{L} |_{\rm fermion} &= 2i \delta \left( \Tr \left[\bpsi^a \bsigma^{\mu}  D_{\mu} \psi_a \right] \right)  \\
&=2i  \Tr \left[ \delta \left(\bpsi^a \right)\bsigma^{\mu}  D_{\mu} \psi_a \right]
+2i \Tr \left[\bpsi^a \bsigma^{\mu}  D_{\mu} \delta \left(\psi_a\right) \right] 
+2i  \Tr \left[\bpsi^a \bsigma^{\mu}   i[\delta \left(A_{\mu}\right),\psi_a] \right] \nonumber\\
&=2i  \Tr \left[ \delta \left(\bpsi^a \right)\bsigma^{\mu}  \partial_{\mu} \psi_a \right]
+2i \Tr \left[\bpsi^a \bsigma^{\mu}  \partial_{\mu} \delta \left(\psi_a\right) \right] 
\end{align}
and in the last line we passed to the $\lambda \to 0$ limit.  

One can next verify that
\begin{align}
\delta \mathcal{L} |_{\rm gauge}
&=    \partial^{\mu} \Tr \left[   \frac{2}{2\sqrt{2}} F_{\mu \nu} \bepsilon^a   \bar{\sigma}^{\nu} \psi_a  \right] -    \Tr \left[   \frac{2}{2\sqrt{2}} \left(D^{\mu} F_{\mu \nu} \right) \bepsilon^a   \bar{\sigma}^{\nu} \psi_a  \right] + \mathrm{h.c.} ,
\label{eq:GaugeVariation}
\end{align}
while
\begin{align}
\delta \mathcal{L} |_{\rm fermion}
&= \partial_{\mu} \Tr \left[  \frac{-1}{2\sqrt{2} } \bepsilon^a \bsigma^{\nu} \sigma^{\alpha}\,  F_{\alpha \nu} \bsigma^{\mu}  \psi_a  + \mathrm{h.c.}\right]  \nonumber \\
&+  \Tr \left[  \frac{2}{2\sqrt{2} } \bepsilon^a \bsigma^{\nu} \partial^{\alpha} F_{\alpha \nu}   \psi_a   \right] +\Tr \left[  \frac{2}{2\sqrt{2} } \bpsi^a
\bsigma^{\nu} \partial^{\alpha} F_{\alpha \nu} \epsilon_a \right] 
\label{eq:FermionVariation}
\end{align}
where we used 
\begin{align}
\bsigma^{\nu} \sigma^{\alpha} \bsigma^{\mu} &= - \eta^{\nu \mu} \bsigma^{\alpha} +  \eta^{\alpha \mu} \bsigma^{\nu}
+\eta^{\nu \alpha} \bsigma^{\mu} + i \epsilon^{\nu \alpha \mu \kappa} \bsigma_{\kappa} .
\label{eq:ThreeSigmaIdentity}
\end{align}
twice.  

So acting on $\mathcal{L}$, the field variations above lead to 
\begin{align}
\delta \mathcal{L}=  \partial_{\mu} G^{\mu}
\end{align} 
where
\begin{align}
G^{\mu} = \Tr \left[   \frac{2}{2\sqrt{2}} F^{\mu \nu} \bepsilon^a   \bar{\sigma}_{\nu} \psi_a \right]+ \Tr \left[  \frac{-1}{2\sqrt{2} } \bepsilon^a \bsigma^{\nu} \sigma^{\alpha}\,  F_{\alpha \nu} \bsigma^{\mu}  \psi_a \right]  + \mathrm{h.c.} ,
\end{align}
so that the variation of the action is a total derivative.  This means that these field variations are associated with a fermionic symmetry, with $N_f$ spin-$3/2$ Noether currents
\begin{align}
J^{\mu  \dot{\kappa}}_a = \frac{1}{\sqrt{2}}\Tr\left[ F_{\rho \nu} (\bsigma^{\nu} \sigma^{\rho} \bsigma^{\mu} \psi_{a})^{\dot{\kappa}}\right]\, .
\end{align}
One can easily verify that these Noether currents are conserved at $\lambda = 0$: $\partial_{\mu} J^{\mu}_a$ vanishes on-shell by using \eqref{eq:EoM}.  Hence there are $4 N_f$ conserved fermionic charges in the $\lambda = 0$ limit of adjoint QCD, in flat space.

While it is amusing that there is a fermionic symmetry in flat-space adjoint QCD at $\lambda = 0$, this observation raises two obvious questions. First, it would be very interesting to work out how this fermionic symmetry behaves on $S^3 \times S^1$ in the limit  $R\Lambda \ll 1$.   To answer this question one would first need to understand the full symmetry algebra generated by the combination of the fermionic charges, the bosonic flavor charges, and the Poincare charges.  An answer to this question with $\lambda = 0$ should already be quite interesting since it seems quite unlikely that such a symmetry algebra could be a standard superalgebra.  Indeed, there are reasons to suspect that the full symmetry algebra may end up being infinite-dimensional.\footnote{We are very grateful to S.~Dubovsky for alerting us to this possibility and for related discussions.} Second, it would be even more interesting to understand whether a generalization of this kind of symmetry can survive at $\lambda \neq 0$ in the large $N$ limit of adjoint QCD.   An exploration of some of these issues is now in progress\cite{SymmetriesFuture}.

\section{Conclusions}
\label{sec:Conclusions}
We have studied adjoint QCD in the large $N$ limit on $S^3_R \times S^1_L$ in the weakly coupled limit $R \Lambda \ll 1$. Despite being weakly coupled, these theories have all of the features one would expect from any well-to-do confining large $N$ theory, with a Hagedorn spectrum of stable hadrons created by single-trace operators. We have found that the bosonic and fermionic density of states have a Hagedorn growth.  Nevertheless, the bosonic and fermionic states appear to be essentially degenerate up to a curvature-driven misalignment for any $N_f \geq1$ as discussed in Sec.~\ref{sec:InstabilitiesAndDisappearance}.   The spatially compactified theory was explicitly shown to have no Hagedorn instabilities due to enormous cancellations between bosons and fermions.  Our analysis shows that adjoint QCD stays in the confining phase persists of any $L$, and hence enjoys large $N$ volume independence for any $L$.  We also found that the difference of bosonic and fermionic Casimir energies vanishes.\footnote{In \cite{Basar:2014hda} it is shown that the sum of the Casimir energies also vanishes at $N=\infty$ in the confined phase, and these two observations taken together imply that these Casimir energies are actually separately zero.}  As discussed in Sec.~\ref{sec:Misaligned} large $N$ adjoint QCD on $S^3 \times S^1$ appears to provide a field theoretic example of the idea of misaligned supersymmetry from string theory.

Our results involve conspiracies between the energies and degeneracies of all of the bosonic and fermionic hadronic excitations.  This is quite surprising, since the family of theories we consider is not supersymmetric at any finite $N$, and so cannot have any fermionic symmetries at finite $N$.   Since our results are obtained in the large $N$ limit, rather than at finite $N$, they cry out for an explanation in terms of an emergent fermionic symmetry  large $N$, as advocated in \cite{Basar:2013sza}.  We have shown that such a symmetry emerges  for $N_f=1$, as discussed in Sec.~\ref{sec:Symmetries}, while for $N_f > 1$ we were only able to make some preliminary observations.

The analysis we have done takes essential advantage of the $R \Lambda \ll 1$ weak-coupling limit, and it is not clear how to generalize it to study the decompactified regime $R \Lambda \gtrsim 1$, where the theory becomes strongly coupled.  Understanding what happens with large $N$ volume independence once $R \Lambda \gtrsim 1$ presumably requires different techniques, such as numerical lattice calculations, or a refined understanding of the large $N$ symmetries of adjoint QCD.

Indeed, the most pressing direction for future work is understanding whether (and if so, how) fermionic symmetries emerge at large $N$ in $N_f>1$ adjoint QCD, either on $S^3 \times S^1$ or directly on $\mathbb{R}^4$.  The stakes are high: historical experience with supersymmetry shows that fermionic symmetries can be very powerful, and given the very close relationship between adjoint QCD and a sensible large $N$ limit of real-world QCD\cite{Armoni:2003gp,Armoni:2004uu}, finding such symmetries in adjoint QCD could be very useful both theoretically and phenomenologically.  

{\bf Acknowledgements. }  We are grateful to O.~Aharony, A.~Armoni, T.~Cohen, S.~Dubovsky, V.~Gobernko, Z.~Komargodski, B.~Lucini, M.~Shifman, and M.~Yamazaki for fun and inspirational discussions at various stages of the long gestation of this paper.  We owe a special thanks to M.~\"Unsal and D.~Dorigoni for encouragement, extensive discussions, and collaboration on related topics.    We are also deeply grateful to K.~Dienes for very inspirational discussions and comments on the manuscript.   This work is supported by the U.S. Department of Energy under the grants DE-FG-88ER40388 (G.~B.), DE-FG02-93ER-40762 (G.~B.), and DE-SC0011842 (A.~C.). 

\appendix
\section{Single particle partition functions}
\label{AppendixLetter}
In this appendix, which is included to make the paper as self-contained as possible, we give the derivation of the standard expressions\cite{Sundborg:1999ue,Polyakov:2001af,Aharony:2003sx} for the free single particle partition functions for scalar, fermion, and Maxwell fields\footnote{There are some typos in the vector partition function in \cite{Sundborg:1999ue}.} on $S^3 \times S^1$.   

The idea of the derivation is to use the conformal symmetry of the $\lambda = 0$ theory to map a state with energy $E$ on $S^3\times \mathbb R$ to a local operator on $\mathbb R^4$ with dimension $\Delta=E$. With this state-operator mapping, the problem boils down to counting operators with a given dimension $\Delta$. These operators are the conformal descendants $Y_{(n)}$ of a given primary field $Y$ satisfying the condition
\begin{align}
Y_{(n)} = \partial_{\alpha_1} \partial_{\alpha_2} \cdots \partial_{\alpha_n} Y \, .\label{family}
\end{align}
For a primary $Y\equiv Y_{(0)}$ with dimension $\Delta_{Y}$,  the scaling dimension of the descendant $Y_{(n)}$ in~\eqref{family} is $\Delta_n=\Delta_Y+n$. Then the single particle partition function associated with $Y$ can be written as
\begin{align}
z_Y(q)=\sum_\Delta d_\Delta q^\Delta=q^{\Delta_Y}\sum_{n=0}^\infty d_n q^n
\end{align} 
where $q=e^{-\beta/R}$.  We now need to compute $d_n$ to determine $z_{Y}(q)$.   In doing this, it is important that the contributions of operators that include the equation of motion, $\mathcal DY=0$, be subtracted from the partition function since 
\begin{align}
Y_{(n)}^{\rm EOM} = \partial_{\alpha_1} \partial_{\alpha_2} \cdots \partial_{\alpha_n} \big( \mathcal D Y \big) = 0 \,.
\end{align}  
For conformally-coupled scalars and fermions, this is the only constraint that must be taken into account in computing the single-particle partition functions, while for Maxwell fields there are additional constraints from gauge invariance, which we discuss separately.

Taking the equation of motion subtraction is easy to do after observing that the degeneracy of the level-$n$ descendant of $\mathcal D Y$ is identical to the degeneracy of the level-$n$ descendants of $Y$ with a shift in the dimension by the mass dimension of the operator $\mathcal{D}$ which defines the equation of motion, $[\mathcal{D}]$. Or, in short, $\Delta(Y_{(n)}^{\rm EOM}) =  [\mathcal{D}] + \Delta_{Y} + n$. Then, the single particle partition function becomes
\begin{align}
z_Y(q)=\,q^{\Delta_Y} (1-q^{[\mathcal D]})\sum_{n=0}^\infty \hat d_n q^n
\end{align}
where $\hat d_n$ counts the number of different operators of the form \eqref{family} without any restriction. The number of different combinations of $\partial_{\alpha_1}\dots\partial_{\alpha_n}$ is ${(n+3)!\over n!\,3!}$.\footnote{In $d$ dimensions one gets ${(n+d-1)!\over n!\,(d-1)!}$.} Labeling the number of internal degrees of freedom of $Y$ as $\mathcal N_{Y}$ we obtain
\begin{align}
\hat d_n=\mathcal N_{Y}{(n+3)!\over n!\,3!}\,.
\end{align} 
Consequently we arrive at the result
\begin{align}
z_Y(q)=\mathcal N_Y {q^{\Delta_Y}(1-q^{[\mathcal D]}) \over (1-q)^4}
\end{align}
This expression holds for fermions and scalars.  Specializing to a conformally-coupled free real scalar $\phi$, we have  $\Delta_\phi=1$, $\mathcal N_{\phi}=1$, and the operator defining the equation of motion is the Laplacian with $[\nabla^2]=2$. Hence
\begin{align}
z_{\phi}(q)={q+q^2\over(1-q)^3}\,.
\end{align}  
For a free Majorana fermion, we set $\Delta_\psi=(d-1)/2=3/2$, and use $\mathcal N_\psi=2^{d/2}=4$. With  $[\slashed{\mathcal D}]=1$ for the Dirac operator, we obtain
\begin{align}
z_{\psi}(q)={4\,q^{3/2}\over(1-q)^3}\,.
\end{align}

For a Maxwell gauge field, in addition to the constraint that follows from equation of motion,  an additional constraint from gauge fixing has to be imposed on the operators. Let us again start with the most general descendant of the gauge field $A_\mu$ which has dimension $n+1$,
\begin{align}
\partial_{\alpha_1} \cdots \partial_{\alpha_n} A_{\mu} \, . \label{A1}
\end{align}
There are $4\frac{(n+3)!}{n!\,3!}$ such operators, where $\mathcal N_{A_\mu}=4$ since there are $4$ components of the gauge field. 
We now fix the gauge and project out the non-gauge-invariant operators. It is convenient to work in the so-called ``radial gauge'' where 
\begin{align}
A_{\alpha_1} = 0 \, , \, \partial_{\alpha_1} A_{\alpha_2} + \partial_{\alpha_2} A_{\alpha_1} = 0 \, , \, \cdots \, , \, \sum_{\rm permutations} \partial_{\alpha_1} \cdots \partial_{\alpha_n} A_{\alpha_{n+1}} = 0 \, . \label{A2}
\end{align}
It is easy to see that these constraints project-out all non-invariant states, at levels $n = 0$ and $n = 1$. For $n = 0$, $A_{\mu}$ is not invariant, and should be projected out. For $n = 1$, there would naively be $16$ descendants of $A_{\mu}$. But the only single-derivative gauge-invariant object is the field-strength tensor, $F_{\mu \nu}$. Subtracting the symmetric combination of derivatives and vector indices in \eqref{A2} from those appearing in \eqref{A1} leaves only the antisymmetric combination, $F_{\mu \nu}$. 

The number of symmetric combinations given in Eq.~\eqref{A2} with dimension $n+1$ is simply ${(n+4)!\over(n+1)!\,3!}$. Therefore, the off-shell vector partition function is 
\begin{align}
z^{\rm off-shell}_{V}(q)=\sum_{n=0}^\infty\left(4\frac{(n+3)!}{n!\,3!}-{(n+4)!\over(n+1)!\,3!}\right)q^{n+1}={4q-1\over(1-q)^4}+1\,.
\label{eq:vos}
\end{align}
We still have to project out the operators nullified by the equation of motion, $\partial_\mu F_{\mu\nu}=0$, from Eq.~\eqref{eq:vos}. This procedure can be carried on in two steps. First, we identify the family of gauge fixed descendants that are nullified by the equation of motion:
\begin{align}
\partial_{\mu} (\partial_\mu A_{\nu} - \partial_{\nu} A_{\mu} ) = 0 \,,\, \cdots \,,\, 
\partial_{\alpha_1} \partial_{\alpha_2} \cdots \partial_{\alpha_n} \partial_{\mu} (\partial_{\mu} A_{\nu} - \partial_{\nu} A_{\mu} )= 0 \,,\, \cdots \,. \label{A4}
\end{align} 
The number of such operators with dimension $n+1$ is $4{(n+1)!\over (n-2)!\,3!}$, which follows from counting the number of symmetric combinations of $n-2$ derivatives and multiplying it by the number of components of $A_\mu$.  However, not all of the constraints in Eq.~\eqref{A4} are independent. Because $F_{\mu \nu}$ is antisymmetric in its two indices, any symmetric contraction one of the extra derivatives hitting the equation of motion in a descendant will identically vanish \emph{independently} of the equation of motion, i.e.
\begin{align}
\partial_{\mu} \partial_{\nu} (F_{\mu \nu}) = 0 \,,\, \cdots \,,\, 
\partial_{\alpha_1} \partial_{\alpha_2} \cdots \partial_{\alpha_n} \partial_{\mu} \partial_{\nu} (F_{\mu \nu}) = 0 \,.\, \cdots \,. \label{A5}
\end{align}
The second step is to \textit{add} these terms back to correct for the double counting. The number of these descendants at level $n+1$ is ${n!\over (n-3)!\,3!}$ which is the number of symmetric combinations of $n-3$ derivatives that  hit $\partial_\mu\partial_\nu F_{\mu\nu}$.
We then find that
\begin{align}
z_V^{\rm EOM}(q) = \sum_{n=1}^\infty \left(4{(n+1)!\over (n-2)!\,3!}-{n!\over (n-3)!\,3!}\right)q^{n+1}=\frac{(4-q) q^3}{(1-q)^4}\,.
\end{align}
Putting everything together, the vector single particle partition function is obtained as
\begin{align}
z_V(q)=z_V^{\rm off-shell}(q)-z_V^{\rm EOM}(q)=\sum_{n=1}^\infty 2 n (n+2 ) q^{n+1}=\frac{2 (3-q) q^2}{(1-q)^3}
\end{align}

\section{Analytic expressions for Hagedorn temperatures}
\label{Ap:Roots}
In this appendix, we give the analytical expression for the roots that encode the singularities of the thermal and twisted partition functions given in Eqs. \eqref{eq:ZNfThermal} and \eqref{eq:ZNfTwisted}. As explained in Section \ref{sec:ThermalInstability}, the closest root to the origin along the real axis controls the Hagedorn growth of the density of states. In the thermal compactification, this closest root, $r^*$, also controls the Hagedorn temperature, $T_H$, via the relation $T_H = -{1\over 2 R \log r_*}$.

For the thermal compactification, the relevant polynomial whose roots encode the singularities for the thermal compactification, given in Eq. \eqref{polynomial}, is $P(Q)=Q^6-3Q^4-4N_fQ^3-3Q^2+1$. 
The $Q\leftrightarrow Q^{-1}$, $T$-reflection symmetry forces the roots to come in reciprocal pairs, which we label as $\{r_i,r_i^{-1}\}$ with $i=1,2,3$. It is also useful to define
\begin{align}
R_i=r_i+{1\over r_i},\quad i=1,2,3\,.
\end{align}
Writing the equation for the roots as
\begin{align}
0=P(Q)=\prod_{i=1}^3(Q^2-R_i Q -1)
\end{align}
leads to the set of equations
\begin{align}
 \sum_{i=1}^3 R_i =0,\qquad \prod_{1\leq i<j\leq 3} R_i R_j=-6,\qquad \prod_{i=1}^3 R_i =4N_f\,.
 \label{eq:Ris}
\end{align}
Solving Eqs. \eqref{eq:Ris} simultaneously, for $N_f\geq2$ , we arrive at the expressions
\begin{align}
r_1&=\frac{\kappa ^2+2-\sqrt{\kappa^4+4}}{2 \kappa} \nonumber \\
r_2&= -{1 \over 16\kappa^2} \left[ \kappa^3+2\kappa -2\sqrt{\eta} +\left((\kappa^3+2\kappa-2\sqrt{\eta})^2-16\kappa^4\right)^{1/2}\right] \nonumber \\
r_3&=-{1 \over 16 \kappa^2} \left[ \kappa^3+2\kappa +2\sqrt{\eta} -\left((\kappa^3+2\kappa+2\sqrt{\eta})^2-16\kappa^4\right)^{1/2}\right]
\label{r123}
\end{align}
where
\begin{align}
\kappa &\equiv \left(2N_f+2\sqrt{N_f^2-2}\right)^{1/3}\\
\eta &\equiv 3\left( \kappa^4-N_f \kappa^3- \kappa^2+  2 \right)\,.
\end{align}
Among these roots and their reciprocals, the one closest to origin along the real axis is $r_1$.

 For $N_f=1$, there is a further simplification. The polynomial $P(Q)$ can be factored as
\begin{align}
P(Q)=(1+Q)^2(1-2Q-2Q^3+Q^4)\,\qquad(N_f=1),
\label{Nf1polynomial}
\end{align}
and has roots 
\begin{align}
r_1&=\frac{1}{2}-\frac{\sqrt{2} \sqrt[4]{3}}{2}+\frac{\sqrt{3}}{2}\nonumber\\ 
 r_2 &= \frac{1}{2}-\frac{i \sqrt[4]{3}}{\sqrt{2}}-\frac{\sqrt{3}}{2}=-e^{i \sin^{-1}(3^{1/4}/\sqrt{2})} \qquad(N_f=1)\nonumber \\
 r_3&=-1\,
 \label{Nf1roots}
\end{align}
with their reciprocals.
For spatial compactification, the leading singularity  is $-r_1$ and it is on the negative real axis. The rest of them can be obtained by substituting $N_f\to-N_f$ in\eqref{r123} and their reciprocals. 

\section{The representation of the twisted partition function in terms of elliptic functions}
\label{Ap:Elliptic}

The $T$- reflection symmetry of the twisted partition function allows one to express it in terms of elliptic functions. To see this, let us start with the infinite product form given in Eq. \eqref{eq:ZNfTwisted}. For such a representation, it is convenient to use the variables $\xi \equiv Q^{1/2}=e^{-{L\over4 R}}\equiv e^{i \pi \tau}$, $r_i\equiv e^{2 i \nu}$, where  $r_i$ and  $r_i^{-1}$ are the roots of $P(Q)$ given in \eqref{polynomial}. The denominator of the twisted partition function can be written as
\begin{align}
\prod_{i=1}^3 \prod_{k=1}^\infty (1 + r_i \xi^{2k})(1 + r_i^{-1} \xi^{2k})&=\left(\prod_{m=1}^\infty{1\over(1-Q^m)^3}\right)\prod_{i=1}^3 \prod_{k=1}^\infty\left[(1+2 \cos(2 \nu_i) \xi^{2k}+\xi^{4k})(1-\xi^{2k})\right]\nonumber\\
&=\left({Q^{{1\over8}}\over \eta^3(\tau)}\right)\prod_{j=1}^3\left[ {\vartheta_2(\nu_j |e^{i\pi\tau})\over 2 \cos(\nu_j) \xi^{1/4}}\right]={Q^{-{1\over4}}\over \eta^3(\tau)}\prod_{j=1}^3 {\vartheta_2(\nu_j |e^{i\pi\tau})\over r_j^{1/2}+r^{-1/2}_j}\,,
\end{align}
where we have used the Jacobi triple product to obtain the theta function. The numerator can also be expressed in terms of the Dedekind eta function,
\begin{align}
\prod_{k=1}^\infty(1-Q^{2k})^3=Q^{-1/4}\eta^3(2\tau)\,.
\end{align}
Putting everything together, we obtain our final result
\begin{align}
\tilde Z_{\rm QCD[Adj]}=\eta^3(2\tau)\eta^3(\tau) \prod_{j=1}^3\left[{r_j^{1/2}+r_j^{-1/2}\over \vartheta_2(\nu_j|Q^{1/2})}\right]\,.
\label{ellipticZ}
\end{align}
Note that the above expression can be simplified further when $N_f=1$. In fact, due to the double root $Q=-1$ for $N_f=1$, the formula \eqref{ellipticZ} should be used with care. Let us analyze this case explicitly. Using the expressions for the roots give in \eqref{Nf1roots}, we can write 
\begin{align}
\tilde Z_{N_f=1}&=\prod_{m=1}^\infty(1-Q^{2m})^3\prod_{i=1}^2\prod_{k=1}^\infty {1\over (1-Q^m)^2 (1+ r_iQ^k)(1+r_i^{-1} Q^k)}= \eta^3(2\tau)\prod_{j=1}^2\left[{r_j^{1/2}+r_j^{-1/2}\over \vartheta_2(\nu_j|Q^{1/2})}\right]\nonumber\\
&={\sqrt{6}\,\eta^3(2\tau)\over \vartheta_2\left({i\over 2} \log \left({1\over2}- {\sqrt[4]{3}\over\sqrt{2}}+{\sqrt{3}\over2}\right)|Q^{1/2}\right)\vartheta_1\left(\frac{1}{2} \sin ^{-1}\left(\frac{\sqrt[4]{3}}{\sqrt{2}}\right)|Q^{1/2}\right)},\quad(N_f=1)
\end{align}
where we used the identity $\vartheta_1(z|e^{i\pi\tau})=-\vartheta_2(z+{\pi\over2}|e^{i\pi\tau})$.

\section{Numerical computation of the twisted Casimir energy}
\label{Ap:Casimir}

\begin{figure}[t]
 \centering
\includegraphics[width=0.75\textwidth]{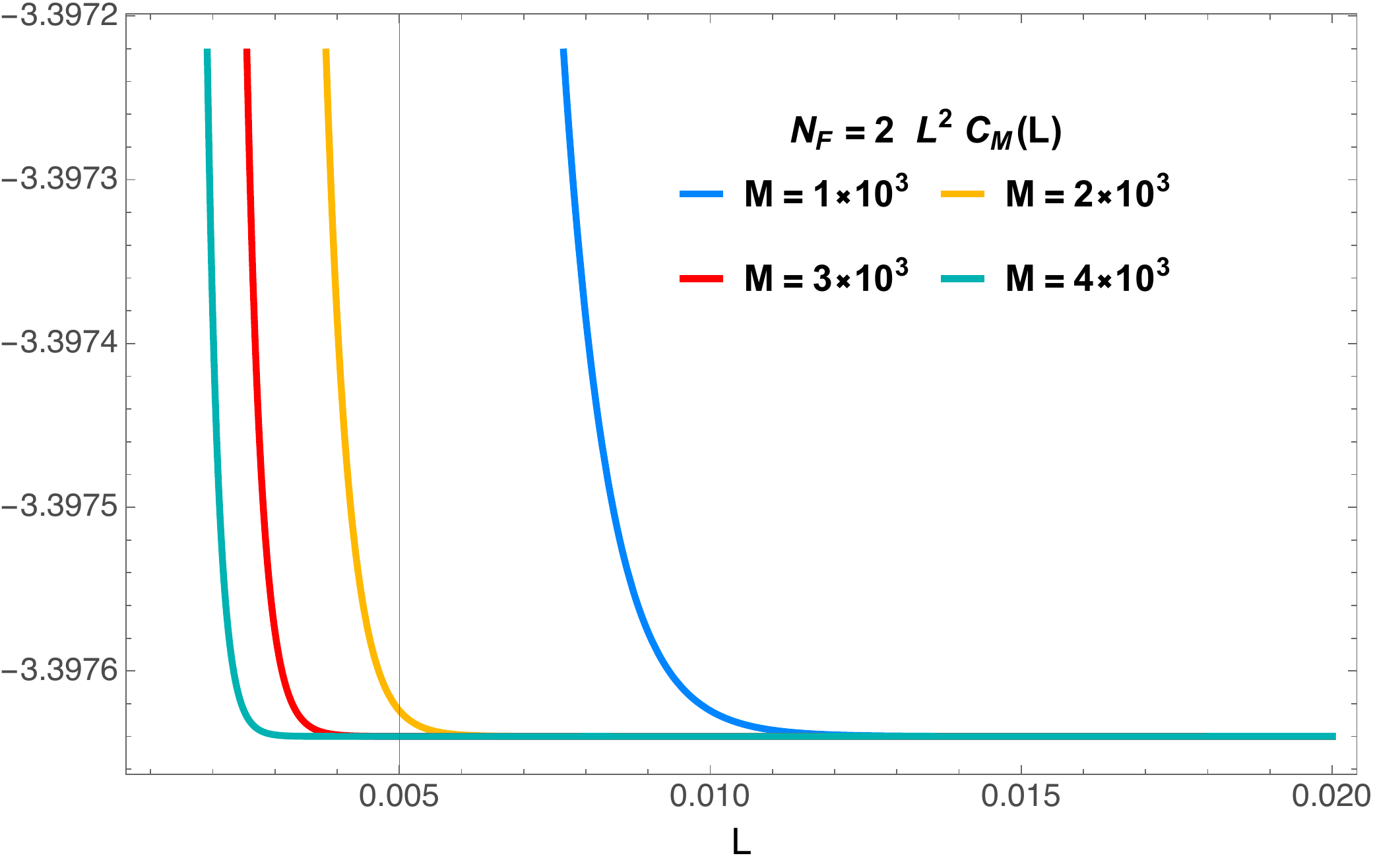} 
\caption{Behavior of $L^2 C(L)$ at small $L$ for $N_f=2$ (as an example) as a function of a cutoff $M$ on the upper end of the sum in \eqref{eq:ZSTrepeated}. }\label{fig:Convergence}
\end{figure}

\begin{table}
\begin{center}
  \begin{tabular}{| c | c | c | c | c|}
    \hline
     \multicolumn{5}{|c|}{$\mathbf{N_f=1}$} \\
     \hline
    M & $C R$ & $C_1$ & $C_2 R$ & $\sigma$ \\ \hline
    $5.00 \times 10^2$ & $2.74 \times 10^{-2} $ & $1.94 \times 10^{-3}$ & $1.12$ &  $2.53 \times 10^{-7}$ \\ 
    $8.00 \times 10^2$ & $4.64 \times 10^{-3} $ & $2.78 \times 10^{-4}$ & $1.12$ &  $6.34 \times 10^{-8}$ \\ 
    $1.50 \times 10^3$ & $1.51 \times 10^{-3} $ & $6.95 \times 10^{-5}$ & $1.12$ &  $2.07 \times 10^{-8}$ \\ 
    $8.00 \times 10^3$ & $1.18 \times 10^{-4} $ & $1.60 \times 10^{-6}$ & $1.12$ &  $3.45 \times 10^{-9}$\\ 
    \hline
  \end{tabular}
  \begin{tabular}{| c | c | c | c | c |}
   \hline
     \multicolumn{5}{|c|}{$\mathbf{N_f=2}$} \\
     \hline
         M & $C R$ & $C_1$ & $C_2 R$ & $\sigma$ \\ \hline
    $5.00 \times 10^2$ & $5.45 \times 10^{-2} $ & $3.88 \times 10^{-3}$ & $3.40$ & $1.68 \times 10^{-7}$  \\ 
    $8.00 \times 10^2$ & $8.76 \times 10^{-3} $ & $5.38 \times 10^{-4}$ & $3.40$ & $4.24 \times 10^{-8}$ \\ 
    $1.50 \times 10^3$ & $2.66 \times 10^{-3} $ & $1.31 \times 10^{-4}$ & $3.40$ & $1.46 \times 10^{-8}$ \\ 
    $8.00 \times 10^3$ & $1.61 \times 10^{-5} $ & $2.59 \times 10^{-7}$ & $3.40$ & $5.89 \times 10^{-10}$ \\ 
    \hline
  \end{tabular}
\end{center}
\caption{Best-fit parameters for the low-$L$ behavior of $C(L)$ as a function of the cutoff $M$.  Note that in both the $N_f=1$ and $N_f=2$ theories the twisted Casimir energy $C$ goes to zero as the cutoff $M$ is removed.  
}
\label{table:Casimir}
\end{table}

We compute $C$ numerically.  If we cut off the infinite sum in \eqref{eq:ZSTrepeated} at some high $n = M$, then $C(L)$ rapidly becomes insensitive to $M$ except at  low $L$.  Accessing lower $L$ requires increasing $M$.  In Fig.~\ref{fig:Convergence} we illustrate the dependence of the low-$\beta$ behavior on the cutoff $M$ in $N_f=2$ adjoint QCD.  From the figure it is clear that the leading small $L$ divergence in $C(L)$ is $\sim 1/L^2$.  One can then verify that the $M$-independent small-$L$ regions of $C(\mu)$ can be modeled to a very high accuracy by a polynomial fit function $F(L)$:
\begin{align}
F(L) = C + \frac{C_1}{L}+\frac{C_2 R}{L^2}
\label{eq:FitFunction}
\end{align}

The parameters $C, C_1, C_2$ are read off from a least-squares fit of the $F(L)$ to $C(L)$ at low $L$ for a variety of values of $M$.  We then take $M \to \infty$ limit.  Our results for $N_f=1$ and $N_f=2$ are summarized in Table~\ref{table:Casimir}.   To characterize the quality of the fits to the function \eqref{eq:FitFunction}, the tables also show the value of 
\begin{align}
\sigma = \frac{1}{n} \sqrt{\sum_{L_i} \left(\frac{C(L_i) - F(L_i)}{C(L_i)}\right)^2}
\end{align}
where $L_i, i=1,2,\ldots, n$ are the set of values of $L$ used to do the fit.   A good fit is characterized by $\sigma \ll 1$, which is true for all the cases we show.
Our results for higher $N_f$ are similar.  We find that the best-fit values of $C$ decrease rapidly toward zero with increasing $M$, and an extrapolation to $M = \infty$ results in $C=0$ for all $N_f \ge 1$.   The same is true for $C_1$, while $C_2$ has a non-zero limit which depends on $N_f$.

\bibliographystyle{apsrev4-1}
\bibliography{super_susy} 

\end{document}